\documentclass[aps,pra,notitlepage,twocolumn,amsmath,amssymb,showpacs,nofootinbib]{revtex4-2}

\usepackage[utf8]{inputenc}
\usepackage{graphicx}
\usepackage{amsmath}
\usepackage{bm}
\usepackage{xcolor}
\usepackage{comment}
\usepackage{dsfont} 
\usepackage{soul}

\newcommand{\be}{\begin{equation}}  
\newcommand{\ee}{\end{equation}}
\newcommand{\ba}{\begin{array}}
\newcommand{\ea}{\end{array}}
\newcommand{\bea}{\begin{eqnarray}}
\newcommand{\eea}{\end{eqnarray}}
\newcommand{\bra}{\langle}
\newcommand{\ket}{\rangle}

\newcommand{\nn}{\nonumber}

\makeatletter  
\newcommand{\fmarki}{\ensuremath{\dagger}}
\newcommand{\fmarkii}{\ensuremath{\ddagger}}
                
\def\@fnsymbol#1{{\ifcase#1\or \fmarki\or \fmarkii \else\@ctrerr\fi}}
\makeatother

\begin{document}

\title{Energy shortcut of $N$-level quantum protocols by optimal control}
\author{C. L. Latune, M. B. Puthuveedu Shebeek, D. Sugny, S. Gu{\'e}rin}
%\email{camille.lombard-latune@u-bourgogne.fr}
\affiliation{Universit{\'e} Bourgogne Europe, CNRS, Laboratoire Interdisciplinaire Carnot de Bourgogne ICB UMR 6303, 21000 Dijon, France}

\begin{abstract}
We introduce an energetically-optimal method inspired from Shortcut-To-Adiabaticity (STA) processes, named Quantum-Optimal-Shortcut-To-Energetics (QOSTE). QOSTE produces the same transformation as STA for a given protocol used in quantum technologies or thermodynamics, but at the lowest possible energy cost. In the general case of a $N$- level quantum system, we derive the QOSTE controls using geometrical and optimal control tools, and show that the minimal energy cost is  determined by the length of the geodesic in the rotating frame given by the original protocol. For long control times, the scaling of the ratio between the two energy costs of STA and QOSTE is quadratic in time. We benchmark our results with the Landau-Zener protocol for qubits and STIRAP for three-level systems. We observe a drastic reduction in energy with respect to standard STA methods. Finally, using gradient-based optimization algorithms and highlighting the emerging trade-off between robustness and energy cost, we design robust QOSTE outperforming STA both in robustness and energy efficiency. 
\end{abstract}

\maketitle

{\it \noindent Introduction}. Control of quantum systems is at the core of quantum applications and quantum technologies~\cite{Glaser15,kochroadmap,past-present-future,altafini2012,dong2010,RMPsugny}. A particularly well-known and effective method for designing the controls is the Shortcut-To-Adiabaticity approach (STA)~\cite{Guery-Odelin19,Stefanatos2020,TORRONTEGUI2013,Duncan2025}. STA is a generic term for various techniques that aim to ensure that the quantum system of interest follows an adiabatic trajectory at an arbitrary speed, obtained for instance by adding counterdiabatic (CD-STA) drivings to the system Hamiltonian. Such techniques, which were first introduced in \cite{Unanyan1997}, later in~\cite{Demirplak03, Demirplak05, Demirplak08, Berry09,prl2013_adolfo} and then in a quantum thermodynamic context \cite{Campo14,Deng13,Beau16,Abah19,Hartmann20,Dann20,Dann2020Jan,Deng18} to avoid quantum friction~\cite{Kosloff02,Feldmann03,Feldmann04}, have gained much importance in adiabatic quantum computing~\cite{Hegade21}, experimental state engineering~\cite{Chen21}, and quantum information processing~\cite{Santos20}, to name a few. STA techniques do not offer a definitive answer for accelerating the dynamics since they require an ansatz typically based on physical considerations.
On the other hand, optimal control theory (OCT) is a general mathematical procedure whose goal is to find time-dependent control parameters while minimizing or maximizing a functional that can be the control time-length like in the quantum Brachistochrone problems \cite{Wang15,Koike2022}, or the control energy ~\cite{liberzon-book,dalessandro-book,kochroadmap,kirk2004optimal, Liu2023}. %\st{The mathematical construction of OCT is based on the Pontryagin's Maximum Principle (PMP) which was established in the late 1950s~}\cite{pontryaginbook,leemarkusbook,Ansel24,Boscain21,bonnard_optimal_2012}. \st{Today, OCT has become a powerful tool to optimize a variety of operations in quantum technologies}~\cite{Glaser15,kochroadmap,dupont2021}.

%STA provides a simple answer to perform a Hamiltonian transformation from a given Hamiltonian $H_i$ to a final Hamiltonian $H_f$ while following the adiabatic trajectory defined by a known protocol $H_0(t)$ where $H_0(0) = H_i$ and $H(t_f) = H_f$. 

The main result of the present Letter is the derivation by OCT of 
%\st{ the geodesic corresponding to the energetically-optimal driving that achieves the transformation for a given duration $t_f$.  This is schematically represented in Fig. \ref{fig:geo} where the adiabatic trajectory, achieved by a CD driving, is much longer than the minimal length of the geodesic, while both trajectories connect the successful boundaries of the transformation. This defines}
a method, which we name Quantum Optimal Shortcut-To-Energetics (QOSTE), realizing the same transformations as CD-STA, but where the additional drive is  determined from the minimization of its energy cost for a given duration. QOSTE targets a shortcut trajectory of minimum energy while CD-STA focuses on the preservation of an adiabatic trajectory.  % \Dom{The energy costs of QOSTE and CD-STA correspond respectively to the lengths of the geodesic and of the adiabatic trajectory, as schematically represented in Fig.~\ref{fig:geo}.} 
Additionally, we address robustness issue of QOSTE against variations in Hamiltonian parameters~\cite{Dridi2020,meri2023,vandamme2017, Carolan23}. It is well known that adiabatic passages feature intrinsic properties of robustness~\cite{Vitanov}. However the energy consumed by CD-STA does not contribute a priori to preserve this robustness.
By combining gradient-based optimization algorithms with physical constraints, we design robust QOSTE that are optimized for robustness and energy costs simultaneously; they are shown to outperform CD-STA.

In the following, we consider a Hamiltonian transformation defined by an initial and a final Hamiltonian, denoted by $H_i$ and $H_f$, respectively, and a given protocol $H_0(t)$ from $H_0(0) = H_i$ to $H(t_f) = H_f$. Adiabatic trajectories are defined from the instantaneous eigenstates of $H_0(t)$ \cite{Born1928}. 
 The protocol $H_0(t)$ may be motivated by some thermodynamic protocols such as quantum Otto ~\cite{Campo14,Abah2019,Hartmann2020,Dann20} or Carnot cycles ~\cite{Dann20,Dann2020Jan}, or by other requirements in quantum annealing processes, qubits resets, or even in adiabatic Grover search algorithm~\cite{Daems2008}.
The two main motivations for following adiabatic trajectories are robustness against experimental uncertainties~\cite{Vitanov} and evading unwanted energy transitions (also referred to as quantum friction in the context of quantum thermodynamics~\cite{Kosloff02,Feldmann03,Feldmann04}). 
However, following adiabatic trajectories requires slowing down the system evolution~\cite{Allahverdyan05,Albash12}, which is not desirable for quantum technologies.

CD-STA (as other STA techniques) offers an alernative. CD-STA is based on the ansatz of preserving the adiabatic trajectories defined by $H_0(t)$, at the cost of an additional counter-diabatic drive $V_\text{CD}(t)$~\cite{Demirplak03, Demirplak05, Demirplak08, Berry09}. It can be expressed (in units with $\hbar=1$) as $V_\text{CD}(t) =i \sum_n \big[|\dot n(t)\ket \bra n(t)| - \bra n(t)|\dot n(t)\ket |n(t)\ket\bra n(t)|\big]$,
%\be\label{eq:HCD}
%V_\text{CD}(t) =i\hbar \sum_n \big[|\dot n(t)\ket \bra n(t)| - \bra n(t)|\dot n(t)\ket |n(t)\ket\bra n(t)|\big],
%\ee
where $|n(t)\ket$ denotes the instantaneous eigenstates of the Hamiltonian $H_0(t)$ of eigenenergies $E_n(t)$, $H_0(t)=\sum_n E_n(t)|n(t)\ket\bra n(t)|$.
The effect of $V_\text{CD}(t)$ is actually to cancel the transitions between the energy levels of $H_0(t)$. However, the additional drive $V_\text{CD}(t)$ comes with an additional energy cost~\cite{Moutinho2023,Gois2024} associated to quantum speed limit~\cite{Campbell2017}, power needed to generate controls~\cite{Zheng2016, Torrontegui17, Tobalina19},  increased work fluctuation~\cite{Funo17,delcampo18}, or  classical entropy production due to control signal generation~\cite{Kiely22}. CD driving process suffers from high energy expenditure since it imposes the dynamics to follow the adiabatic trajectory \textit{at all times}. This strong constraint is a priori unnecessary since only the initial and the final states of the protocol matter to achieve the desired transformation. 
The question of energy efficiency in quantum control~\cite{Carolan22} is indeed becoming increasingly important given the intense debate surrounding the energy costs of quantum technologies~\cite{Auffeves22,Asiani23}.

{\it \noindent Energetics}. The energy cost of a control $V(t)$ can be estimated as ~\cite{Kiely22}  
\begin{equation}
{\cal C}[V] = \int_0^{t_f} du ||V(u)||^2,
\end{equation}
where $||V|| := \sqrt{{\rm Tr}[V^\dag V]}$ stands for the Frobenius norm of the CD (QOSTE) driving, $V=V_\text{CD}$ ($V=V_\text{QOSTE}$ denoted in the following as $V_Q$). %$t_f$ denotes the final time of the protocol,
%It is normalized by $\omega_i$, the typical energy scale of the initial Hamiltonian $H_i$. 
 This energy cost relates to the power consumed by the device control which is expected to represent a significant portion of the overall energy bill for quantum technology operations \cite{Asiani23}, and is therefore usually much higher than the cost of the Hamiltonian transformation, which occurs at the quantum level.
Furthermore, applying unnecessarily high energy control to the quantum system can generate extra dissipation and heating, leading to additional energy costs for cooling the experimental setup~\cite{Asiani23}.

\begin{figure}
\begin{center}
  % (a)\includegraphics[width=0.25\textwidth]{H_LZ.png}  \\
  \includegraphics[width=0.3\textwidth]{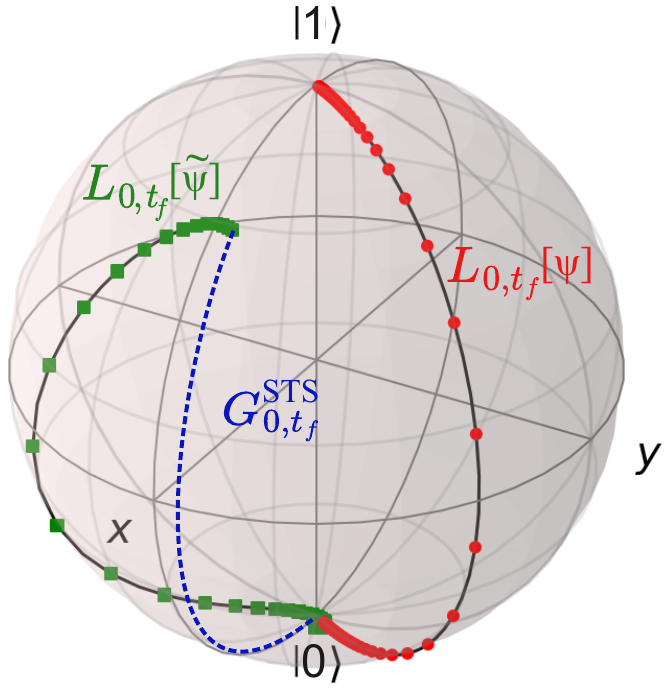}
\end{center}
\caption{Representation of trajectories for a qubit on the Bloch sphere. For a given (unspecified) $H_0(t)$, the red curve represents the trajectory of  $|\psi(t)\ket$ generated by $H_0(t) + V_{CD}(t)$, while the green curve represents the trajectory of $|\tilde\psi(t)\ket = U_0^\dag(t)|\psi(t)\ket$ in the rotating frame with respect to $H_0(t)$. The corresponding geodesic is plotted in blue dashed line.
%The red and green trajectories correspond respectively to the states $|\psi(t)\ket$ and to $|\tilde\psi(t)\ket = U_0^\dag(t)|\psi(t)\ket$ in the rotating frame given by $H_0(t)$, and the corresponding geodesic is plotted in blue dashed line. In the original frame, the goal of the control is to transfer the system from the excited state $|1\rangle$ to the ground state $|0\rangle$. 
}\label{fig:geo}
\end{figure}

In the general case of a $N$- level quantum system, we show below  that the associated counterdiabatic drive $V_{\textrm{CD}}(t)$ implies an energy cost for a unitary transformation that is lower bounded by $\frac{1}{t_f} L_{0,t_f}^2[\tilde U_{\textrm{CD}}]$, where $t_f$ is the time length of the operation and $ L_{0,t_f}[\tilde U_{\textrm{CD}}]=\int_0^{t_f}\sqrt{\textrm{Tr}[\dot{\tilde U}_\text{CD}^\dagger \dot{\tilde U}_\text{CD}]}$ is the length of the adiabatic path in the rotating picture with respect to $H_0(t)$. We use here the evolution operators $\tilde U_\text{CD}(t) := U_0^\dag(t_f) U_\text{CD}(t_f)$, with $U_\text{CD}(t_f):= {\cal T}e^{-i\int_0^{t_f}du [H_0(u) + V_\text{CD}(u)]}$ and  $U_0(t_f):= {\cal T}e^{-i\int_0^{t_f}du H_0(u)}$, ${\cal T}$ denoting the time-ordering operator. The length $ L_{0,t_f}[\tilde U_{\textrm{CD}}]$ of the adiabatic trajectory being larger than a quantity independent of $t_f$ (see Supplementary Material (SM)), the lower bound on the energy cost diverges for fast operations, raising the issue of designing a more energy-efficient method. By contrast, we show that QOSTE has an energy cost determined by $G_{0,t_f}$, the length of the geodesic in the rotating picture with respect to $H_0(t)$, which is always smaller than $ L_{0,t_f}[\tilde U_{\textrm{CD}}]$. A schematic of the different trajectories for a qubit is given in Fig.~\ref{fig:geo}.
 
{\it \noindent The optimization problem.} 
We are looking for controls which generate the same unitary transformation as the CD drive, namely the adiabatic transformation $U_f := \sum_n e^{-i\phi_n(t)} |n(t)\ket\bra n(0)|$, where $\phi_n(t) = \int_0^t du [E_n(u) + i\bra \dot n(u)|n(u)\ket] $ is the sum of the dynamical phase and geometrical phase (the Berry phase)~\cite{Albash12}. More precisely, as for CD-STA, the goal is to find an operator $V(t)$ such that $H(t) = H_0(t) + V(t)$ generates $U_f$, while minimizing the energy cost ${\cal C}[V]$. Since ${\rm Tr}[V_{\textrm{CD}}(t)] = 0$, we assume that ${\rm Tr}[V(t)] = 0$ and therefore we use a basis of traceless Hermitian operators. We can consider for instance the generalized Pauli matrices, $\sigma_k$, $1\leq k\leq N^2-1$ with $\textrm{Tr}[\sigma_k\sigma_\ell]=\delta_{k\ell}$~\cite{schirmer:2004}. The operator $V(t)$ can be spanned in such a basis, $V=\sum_kv_k(t)\sigma_k$, where the $v_k(t)$ are the real control parameters. Here we assume that the control has $N^2-1$ components, i.e. the dimension of the Lie algebra $su(N)$ so any trajectory can be followed by the system, up to a global phase. Then, the energy cost becomes $\mathcal{C}[V]=\sum_k\int_0^{t_f}v_k(t)^2dt$.

For convenience, we use the rotating frame with respect to $H_0(t)$, yielding  $\tilde{V}(t)=\sum_kv_k(t)\tilde{\sigma}_k(t)$ where $\{\tilde{\sigma}_k(t)\}$ is a new basis of traceless Hermitian operators, with $\tilde{\sigma}_k(t)=U_0^\dagger(t) \sigma_k U_0(t)$, $\textrm{Tr}[\tilde{\sigma}_k(t)\tilde{\sigma}_\ell(t)]=\delta_{k\ell}$. The expression of the energy cost is unchanged and the target transformation becomes $\tilde U_f:= U_0^\dagger(t_f)U_f$. Still in the rotating frame, the length of the trajectory described by $\tilde U(t):= {\cal T}e^{-i\int_0^t du V(u)}$ in the space $SU(N)$ is given by
$$
{L}_{0,t_f}[\tilde U]=\int_0^{t_f}\sqrt{\textrm{Tr}[\dot{\tilde{U}}(t)^\dagger \dot{\tilde{U}}(t)]}dt=\int_0^{t_f}\sqrt{\sum_k v_k^2(t)}dt,
$$
with
$$
\textrm{Tr}[\dot{\tilde{U}}^\dagger(t) \dot{\tilde{U}}(t)]=\textrm{Tr}[V^2(t)]=\sum_{k,\ell}v_k(t)v_\ell(t)\textrm{Tr}[\tilde{\sigma}_k(t)\tilde{\sigma}_\ell(t)].$$
Using the Cauchy-Schawrz inequality $[\int_0^{t_f}g_1(t)g_2(t)dt]^2\leq \int_0^{t_f}g_1(t)^2dt\int_0^{t_f}g_2(t)^2dt$, for $g_1(t)=1$ and $g_2(t)=\sqrt{\sum_kv_k^2(t)}$, we arrive at ${L}_{0,t_f}^2[\tilde U]\leq t_f \mathcal{C}[V]$, where the equality holds if and only if $\sum_k v_k^2(t)=\textrm{cst}$~\cite{Boscain21}. 

One can show (see SM) that the operator 
\be\label{eq:Vqoste}
V_{Q}(t) = t_f^{-1} U_0(t) \chi U_0^\dag(t),
\ee
with $\chi$ an Hermitian operator such that $e^{-i\chi} = \tilde U_f$, generates the target unitary $U_f$~\cite{Latune21}. Additionally, $V_{Q}(t)$ is of constant norm, leading to $L_{0,t_f}[\tilde U_{Q}] = t_f {\cal C}[V_{Q}]$. Finally, $\tilde U_{Q}(t)$, the unitary transformation generated by $\tilde V_Q = t_f^{-1}\chi$, follows the geodesic from $\mathbb{I}$ to $\tilde{U}_f$, so that $L_{0,t_f}[\tilde U_{Q}] = G_{0,t_f}$, the length of the geodesic in the rotating frame. We deduce that  $V_{Q}(t)$ defined in Eq.~\eqref{eq:Vqoste} is the control solution minimizing the energy cost. The above geometrical arguments are confirmed by an optimal control analysis using the Pontryagin Maximal Principle (PMP)~\cite{pontryaginbook,leemarkusbook,Ansel24,Boscain21,bonnard_optimal_2012}, see SM.
%We deduce that the optimal solution consists in following the geodesic $\tilde{G}$ in $SU(N)$ from $\mathbb{I}$ to $\tilde{U}_f$. The optimal solution is a control of constant modulus which saturates the bound, $\mathcal{C}[V_{QOSTE}]t_f=\tilde{G}_{0,t_f}^2$.

The same construction can be used for the CD drive, leading to ${L}_{0,t_f}^2[\tilde U_\text{CD}]\leq t_f \mathcal{C}[V_\text{CD}]$. Note that the modulus of $V_\text{CD}(t)$ is not constant so that the inequality is not saturated. Then, since by definition $L_{0,t_f}[\tilde U_\text{CD}]$ is larger than the length of the geodesic $ G_{0,t_f}$, we have the following relation
$$
\mathcal{C}[V_{\text{CD}}]\geq \frac{{L}_{0,t_f}^2[\tilde U_\text{CD}]}{t_f}\geq \frac{{G}_{0,t_f}^2}{t_f}=\mathcal{C}[V_{\text{Q}}],
$$
which provides a simple geometrical comparison of the two energy costs. In addition, the time scaling of their ratio can be estimated for long control times $t_f$ as (see SM)
\be\label{eqratio}
{\cal C}[V_{\text{CD}}]/{\cal C}[V_\text{Q}] \underset{t_f\rightarrow +\infty}{\propto} t_f^2.
\ee
This divergence can be qualitatively explained by the fact that in this limit ${G}_{0,t_f}$ goes to zero whereas ${L}_{0,t_f}[\tilde U_\text{CD}]$ is lower bounded by a positive constant independent of $t_f$.

Finally, if we only need to bring the system to a target state $|\psi_f\ket$, which we refer to as state-to-state transfer (STS), from an initial state $|\psi_0\ket$ to $|\psi_f\ket$, then the control can be much less energetically demanding. This is because for STS, it is not necessary to control the trajectories of all basis states simultaneously as for a unitary transformation. The lowest energy cost of STS corresponds to the energy cost of the cheapest unitary among the ensemble of unitaries realizing the state transfer $|\psi_0\ket$ to $|\psi_f\ket$. We determine (See SM) with OCT the corresponding control with minimal energy cost, denoted $\text{QOSTE}^{\text{STS}}$. It is given by
\be\label{eq:vqsts}
V_\text{Q}^{\text{STS}}(t) = -i\frac{\alpha_f}{t_f}U_0(t)\left( |\psi_0\ket\bra \psi_\perp| - |\psi_\perp\ket\bra \psi_0|\right)U_0^\dag(t), 
\ee
where $\alpha_f=G^{\text{STS}}_{0,t_f}:=\arccos|\bra\psi_0|\tilde\psi_f\ket|$ is the length of the geodesic between the initial state and the target state in the rotating picture $|\tilde \psi_f\ket = U_0^\dag(t_f)|\psi_f\ket$, and $|\psi_\perp\ket$ is the component of $|\tilde\psi_f\ket$ orthogonal to $|\psi_0\ket$.
The energy cost of QOSTE$^\text{STS}$ is simply ${\cal C}[V_\text{Q}^{\text{STS}}] = 2 [G^{\text{STS}}_{0,t_f}]^2/t_f$ (note that the factor 2 does not appear in the relation for unitary transformations), and can be significantly smaller than ${\cal C}[V_Q]$ as shown in the following. 
%where $G^{\text{STS}}_{0,t_f}$ is the length of the geodesic between $|\psi_0\ket$ and $|\tilde \psi_f\ket$, the target state in the rotating picture.

{\it \noindent Landau-Zener model.} As first  benchmark example, we consider the traditional Landau-Zener model, which corresponds to a protocol realizing an adiabatic state-to-state transfer from the ground state $|0\ket$ to the excited state $|1\ket$ of a qubit. The Landau-Zener protocol realizes this transfer with the following control process ~\cite{bason2012,hegerfeldt2013,zenesini2009,trayebirad2010,Duncan2025},
\be
H_0(t) = \Delta(t) \sigma_z + \omega \sigma_x,
\ee
in the limit of inifinitely slow operation with $\Delta(t=\infty) = - \Delta(t=-\infty) = +\infty$. In  realistic experimental conditions, the final time $t_f$ is finite as well as the amplitude of $\Delta(t)$, leading to imperfect STS. To circumvent this limitation, the counterdiabatic drive given by $V_{\text{CD}}(t) = -\tfrac{\dot \Delta(t)\omega}{2(\Delta^2(t) + \omega^2)}\sigma_y$ can be used~\cite{Duncan2025}. Here, the adiabatic transformation $U_f$ realized by $V_\text{CD}(t)$ is $U_f = e^{-i\phi_e}|e_f\ket\bra e_i| + e^{-i\phi_g}|g_f\ket\bra g_i|$, where $|e_i\ket$, $|g_i\ket$, $|e_f\ket$, $|g_f\ket$ are the initial and final energy eigenbases of $H_0(0)$ and $H_0(t_f)$ respectively, and $(|g(t)\rangle,|e(t)\rangle)$ the instantaneous basis at time $t$. 
%can be expressed as $|e_l\ket = \cos\tfrac{\theta_l}{2} |1\ket + \sin\tfrac{\theta_l}{2}|0\ket$ and $|g_l\ket = -\sin\tfrac{\theta_l}{2} |1\ket + \cos\tfrac{\theta_l}{2}|0\ket$, $l=i,f$, with $\theta_l = \arctan (\tfrac{\omega}{\Delta_l}) +\tfrac{\pi}{2}[1-{\rm sign} \Delta_l]$ and $\Delta_i:= \Delta(0)$, $\Delta_f:=\Delta(t_f)$. 
The phases are simply given by $\phi_e = \int_0^{t_f} dt E_e(t)$ and  $\phi_g = \int_0^{t_f} dt E_g(t)$ because $\bra \dot e(t)|e(t)\ket =\bra \dot g(t)|g(t)\ket = 0$ for all times. Here we have $E_e(t) = - E_g(t) = \sqrt{\Delta^2(t) + \omega^2}$. For simplicity, we choose a linear driving function of the form $\Delta(t) = \Delta_0 + \Delta_d\frac{t}{t_f}$.

The QOSTE method for carrying out the transformation $U_f$ is given by the solution~\eqref{eq:Vqoste}, while the QOSTE$^\text{STS}$ is provided by Eq.~\eqref{eq:vqsts} with $|\psi_0\ket = |e_i\ket$ and $|\psi_f\ket = |e_f\ket$. For the settings $\Delta_0/\omega = -10$, $\Delta_d/\omega = 20$ and  $\omega t_f= 1$, the resulting energy costs are ${\cal C}[V_\text{CD}] /\omega= 15.7$, ${\cal C}[V_\text{Q}]/\omega = 7.1$, and ${\cal C}[V_\text{Q}^{\text{STS}}]/\omega = 2.8 $. 

\begin{figure}
\begin{center}
  % (a)\includegraphics[width=0.25\textwidth]{H_LZ.png}  \\
  \includegraphics[width=0.45\textwidth]{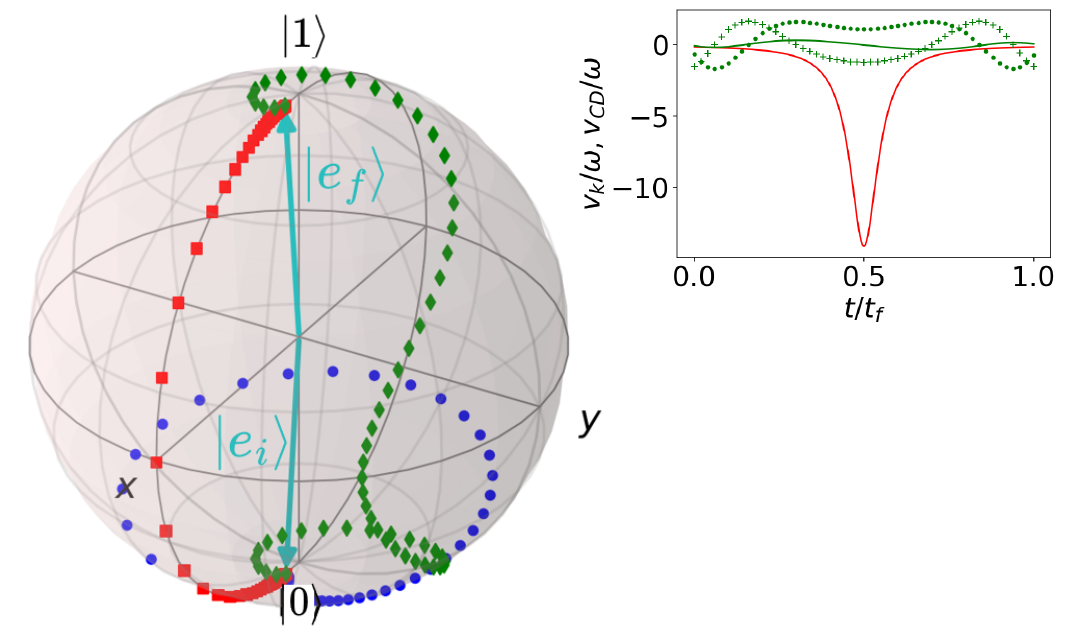}
\end{center}
\caption{%Trajectories (a) of the excited eigenstates of the Hamiltonians $H_0(t)$ (in blue), $H_{\text{CD}}(t) = H_0(t) + V_{\text{CD}}$ (in red) and by $H(t) = H_0(t) + V_\text{opt}(t)$ (in green); (b)
Trajectories of the qubit state  when driven respectively by $H_{\text{CD}}(t) = H_0(t) + V_{\text{CD}}(t)$ (in red), by $H_{\text{QSTS}}(t) = H_0(t) + V_\text{Q}^{\text{STS}}(t)$ (in green), and by $H_0(t)$ (in blue). The goal of the control process is to bring the system from $|e_i\ket$ to $|e_f\ket$. {\bf Inset:} Amplitudes of the control functions for the  QOSTE$^\text{STS}$, $v_k(t) = \frac{1}{\sqrt{2}}{\rm Tr}[V_\text{QOSTE}(t)\sigma_k]$, $k=x$ (green dots), $k=y$ (green pluses), $k=z$ (green solid line), and for the counter-diabatic drive, $v_\text{CD}(t):= \frac{1}{\sqrt2}{\rm Tr}[V_\text{CD}(t)\sigma_y]$ (red solid line). We use $\Delta(t) = \Delta_0 +\Delta_d t/t_f$ with $\Delta_0/\omega = -10$, $\Delta_d/\omega = 20$ and  $\omega t_f= 1$.}\label{fig:LZ}
\end{figure}

 Figure~\ref{fig:LZ} represents the trajectories of the qubit state when driven respectively by $H_0(t)$, $H_{\text{CD}}(t) = H_0(t) + V_{\text{CD}}$ and by $H_\text{Q}^{\text{STS}}(t) = H_0(t) + V_\text{Q}^{\text{STS}}(t)$, for a linear driving function of the form $\Delta(t) = \Delta_0 + \Delta_d\frac{t}{t_f}$. We can verify that the trajectories resulting from the respective CD-STA and QOSTE$^\text{STS}$ methods perform the expected transformation, from $|e_i\ket$ to $|e_f\ket$, whereas the bare trajectory induced by $H_0(t)$ is completely different. This confirms that our choice of final time $t_f$ corresponds to a highly non-adiabatic situation. Indeed, to be adiabatic at all times~\cite{Allahverdyan05,Albash12}, one should choose  $\omega t_f \gg 20$, whereas in Fig.~\ref{fig:LZ} we set $\omega t_f =1$. 

Note that the QOSTE$^\text{STS}$ has an additional advantage, i.e. the control amplitudes are much smaller than the CD drive (see inset of Fig.~\ref{fig:LZ}).
Additionally, for other choices of parameters, and in particular for large $t_f$, we can have arbitrary large reduction of energy cost. However, we mention that in these cases, although the QOSTE$^\text{STS}$ is much more energetically efficient, it tends to be less robust than the CD drive. We explore this emerging relationship between robustness and pulse energy below.

{\it \noindent Robustness.} It is of practical importance, on top of being energetically-optimal, to have protocols that are robust against experimental uncertainties in Hamiltonian parameters or in control parameters (also called control inhomogeneities~\cite{kobzar2004exploring,kobzar2012exploring}). We find similar level of robustness between the CD drive and the QOSTE (see SM). 

\begin{figure}
\begin{center}
  \includegraphics[width=0.47\textwidth]{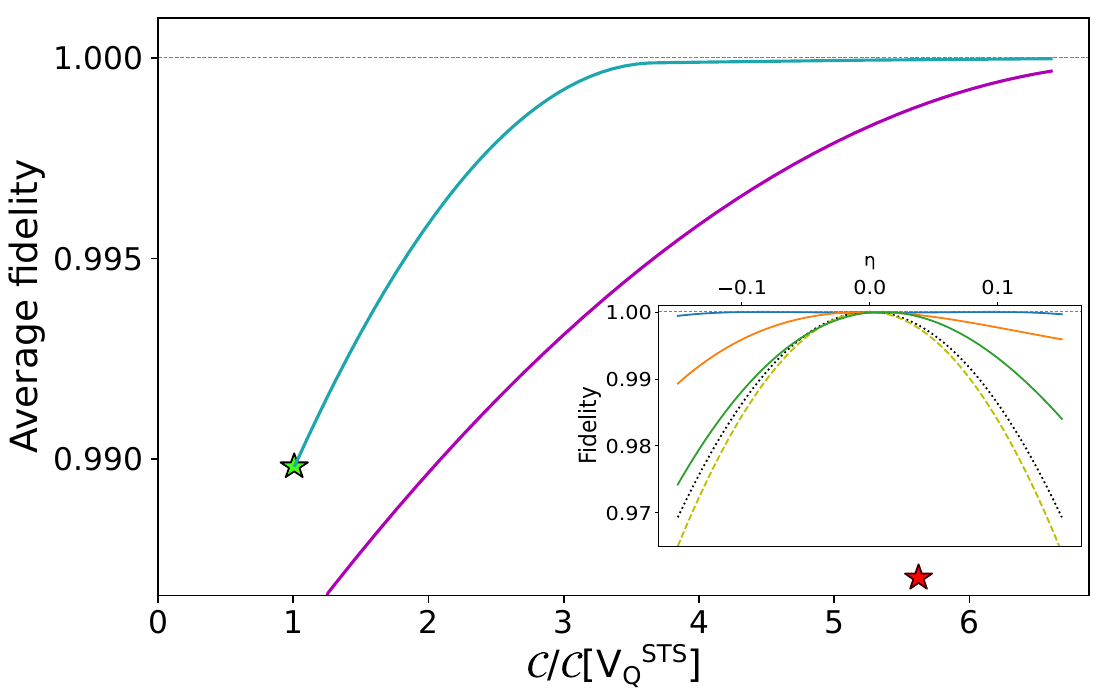}
\end{center}
\caption{ Plots of the average fidelity $\bar{F}_\text{STS}$ (light blue) and $\bar{F}_\text{UT}$ (purple) against the energy cost ${\cal C} = \int_0^{t_f} du ||V(u)||^2$ normalized by ${\cal C}[V_\text{Q}^{STS}]$. The uncertainty range is $\epsilon = 0.15$, and  $N_\eta = 7$. The green star indicates the average fidelity of QOSTE$^\text{STS}$, ${\bar F}_{STS} \simeq 0.989$, while the red star corresponds to the one of CD-STA, ${\bar F}_\text{STS}\simeq 0.987$. {\it Inset:} Plot of the fidelity with respect to the target state $|\bra e_f|\psi_\eta(t_f)\ket|^2$ as a function of the level of uncertainty represented by $\eta$, for the CD drive (in yellow dashed), for QOSTE$^\text{STS}$ (in black dotted) and finally for the GRAPE-designed robust QOSTE$^\text{STS}$ for the respective energy costs ${\cal C}/{\cal C}[V_\text{Q}^{STS}] =1.44 $ (in green), ${\cal C}/{\cal C}[V_\text{Q}^{STS}] =2.39$ (in orange ), and ${\cal C}/{\cal C}[V_\text{Q}^{STS}] = 4.03$ (in blue). }\label{fig:rob_vs_en_inset}
\end{figure}

In the following, we design a control process for STS transfer and for unitary transformation that is energetically optimized and much more robust than the CD drive by using a numerical gradient-ascent method, namely the GRAPE algorithm~\cite{khaneja_optimal_2005,Ansel24} (see also SM). 
We focus on robustness with respect to static uncertainties in the control amplitude, which means that the qubit is driven by an Hamiltonian of the form, 
$$
H_\eta(t)=H_0(t)+(1+\eta)\omega_i\vec{v}\cdot\vec{\sigma},
$$
where $\eta$ is unknown, but belongs to a given range depending on experimental setup. Note that other systematic errors could be treated along the same lines. To design robust controls, we consider an ensemble of $N_\eta$ discrete values of the parameter $\eta$, spanning a range of uncertainty $[-\epsilon,\epsilon]$ (we verify that the robust optimal control does not change for sufficiently large values of $N_\eta$). The aim is then to find controls $v_k$ which generate the target unitary or bring the system to the target state for the $N_\eta$ values of $\eta$, while still minimizing the energy cost. For STS, this is achieved by maximizing the average fidelity over the different values of $\eta$, $\bar F_\text{STS}=\frac{1}{N_\eta}\sum_\eta |\langle e_f|\psi_\eta(t_f)\rangle|^2$, where $|\psi_\eta(t)\rangle$ is the solution of the Schr\"odinger equation for $H_\eta$(t). We then apply a GRAPE algorithm to maximize $\bar F_\text{STS}$ with a fixed energy cost.
We first choose an energy cost equal to ${\cal C}[V_\text{Q}^{STS}]/\omega = 2.8$ and we determine numerically the maximal length $\epsilon$ of the interval such that the controls designed by GRAPE feature the same robustness (average fidelity) as the original QOSTE$^\text{STS}$. We obtain $\epsilon=0.15$. 
%\Cami{Mais si on augmente $\epsilon$, peut-on vraiment obtenir avec GRAPE un contrôle qui est réellement plus robuste que QOSTE tout en ayant le même coût energétique?} %\st{As expected,}\Dom{non, cela depend de l'interval choisi. Si on prend 0.3 au lieu de 0.15, cela n'est pas vrai} \Ste{\st{The protocol we obtain with GRAPE has the same robustness (average fidelity) as QOSTE}}.
Then, we consider a slightly larger initial energy cost, and we derive via GRAPE controls producing a slightly better robustness (still for $\epsilon = 0.15$). Going further, as the energy cost increases, so does the robustness, until tending to a fidelity of one. The results of the numerical optimization are given in Fig.~\ref{fig:rob_vs_en_inset}. The aforementioned trade-off between robustness and energy cost is clearly visible. For completeness, we also design robust QOSTE for unitary transformation with the same procedure by defining an average fidelity over the chosen $N_\eta$ values of $\eta$, $\bar F_\text{UT} = \frac{1}{N_\eta}\sum_\eta |{\rm Tr}[U_f^\dag U_\eta(t_f)]|^2$, where $U_\eta(t_f)$ corresponds to the unitary transformation generated by $H_\eta(t)$.
%presented below. In Fig.~\ref{fig:rob_vs_eta}, we plot the distance to the target state, expressed as $|\bra e_f|\psi_\eta(t_f)\ket|^2$, as a function of the level of uncertainty represented by $\eta$, for several values of $\alpha$ as well as for the energetically optimized protocol without robustness optimization, and the counter-diabatic drive. One clearly sees the large improvement in robustness, especially for small values of $\alpha$. 

% 
% Figure~\ref{fig:rob_en_vs_a} describes the average distance to the target state, $\frac{1}{N_\eta}\sum_\eta |\bra e_f|\psi_\eta(t_f)\ket|^2$ and the energy cost associated with the corresponding control, ${\cal C} = \frac{\omega_i}{2}\int_0^{t_f} dt \vec v^2$ as a function of $\alpha$. These plots highlight that robustness does not come for free and actually requires more energy investment. %This is not really surprising since nothing comes for free in thermodynamics and physics.
%To highlight the energetic cost of robustness even more, Fig.~\ref{fig:en_vs_fid} shows the achieved average fidelity against the associated energetic cost. 
%The trend of high energy costs required for high average fidelity and robustness is clearly visible.
%\Dom{As could be expected, we observe that relatively high energy costs are required for high average fidelity and robustness.}
%To highlight even more the energetic cost of the robustness, Fig. \ref{fig:en_vs_fid} represents the achieved average fidelity against the associated energetic cost. One can clearly see the trend of very large energy costs required for high average fidelity. 
Additionally, it can be seen in Fig.~\ref{fig:rob_vs_en_inset} that for the same energetic cost as for the CD drive (${\cal C}[V_{CD}]/\omega = 15.7$), the GRAPE-designed robust QOSTE is much more robust (more than one order of magnitude). Note also that the energy cost for controlling the relative phase associated with the trajectory of a state orthonormal to $|\psi(0)\ket$ manifests itself as the distance between the curves of $\bar F_\text{STS}$ and $\bar F_\text{UT}$.

{\it \noindent Stimulated Raman Adiabatic passage, STIRAP.} As a second illustration, we consider the optimization of a STIRAP process in a three-level quantum system in a $\Lambda$-configuration~\cite{Vitanov}. In a given frame, the Hamiltonian $H_0(t)$ of the system can be written in the basis $\{|1\ket,|2\ket,|3\ket\}$ as $H_0(t)=\frac{1}{2}\begin{pmatrix}
0 & \Omega_p(t) & 0\\
\Omega_p(t) & 2\Delta & \Omega_s(t)\\
0 & \Omega_s(t) & 0
\end{pmatrix},$ where $\Omega_p(t)$ and $\Omega_s(t)$ are respectively the Pump and Stokes pulses used in a counter-intuitive configuration~\cite{Vitanov,Guery-Odelin19,Liu2023}, and $\Delta$ the detuning with respect to the second energy level. In the adiabatic regime, this control sequence brings the system from the state $|1\rangle$ to the state $|3\rangle$. In the numerical example, we consider the following pulses $\Omega_p(t)=\Omega_0f(t-\tau)$ and $\Omega_s(t)=\Omega_0 f(t)$ with $f(t)=\sin^4(\pi t/T)$, $0\leq t\leq T$, and $f(t) = 0$ otherwise ~\cite{Chen:2010}.

\begin{figure}
\begin{center}
 (a) \includegraphics[width=0.46\textwidth]{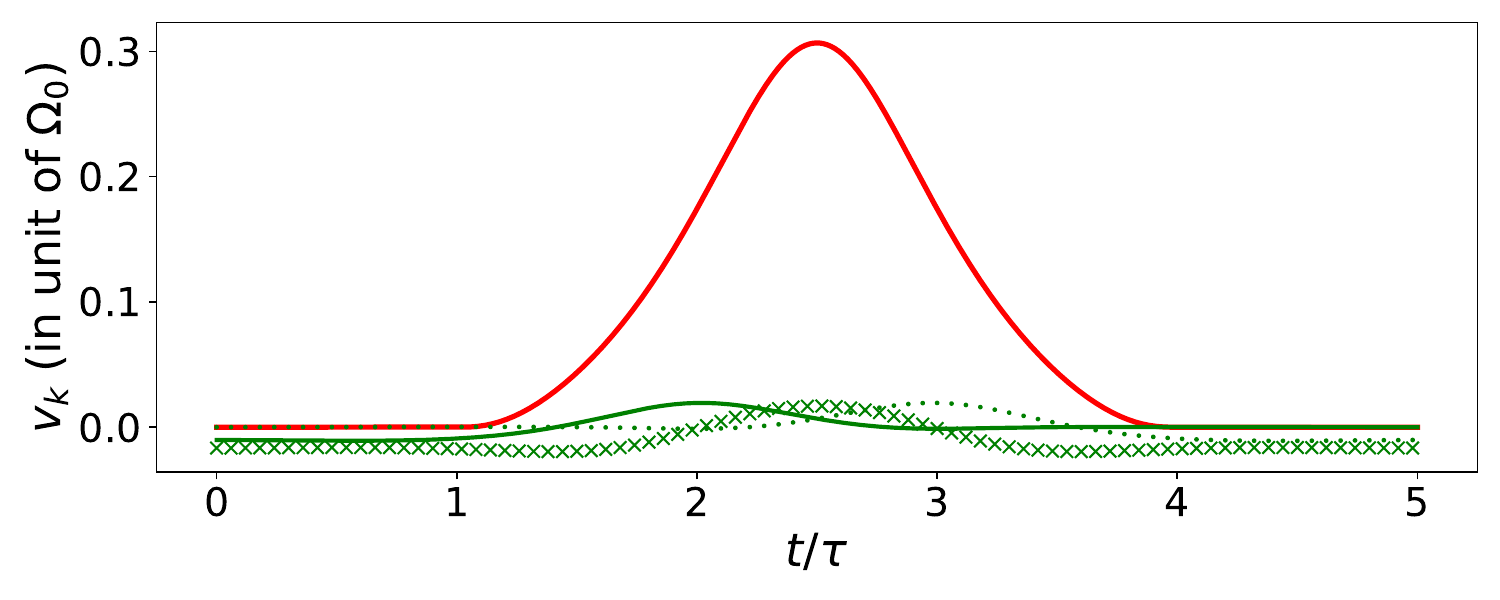}\\
 (b) \includegraphics[width=0.46\textwidth]{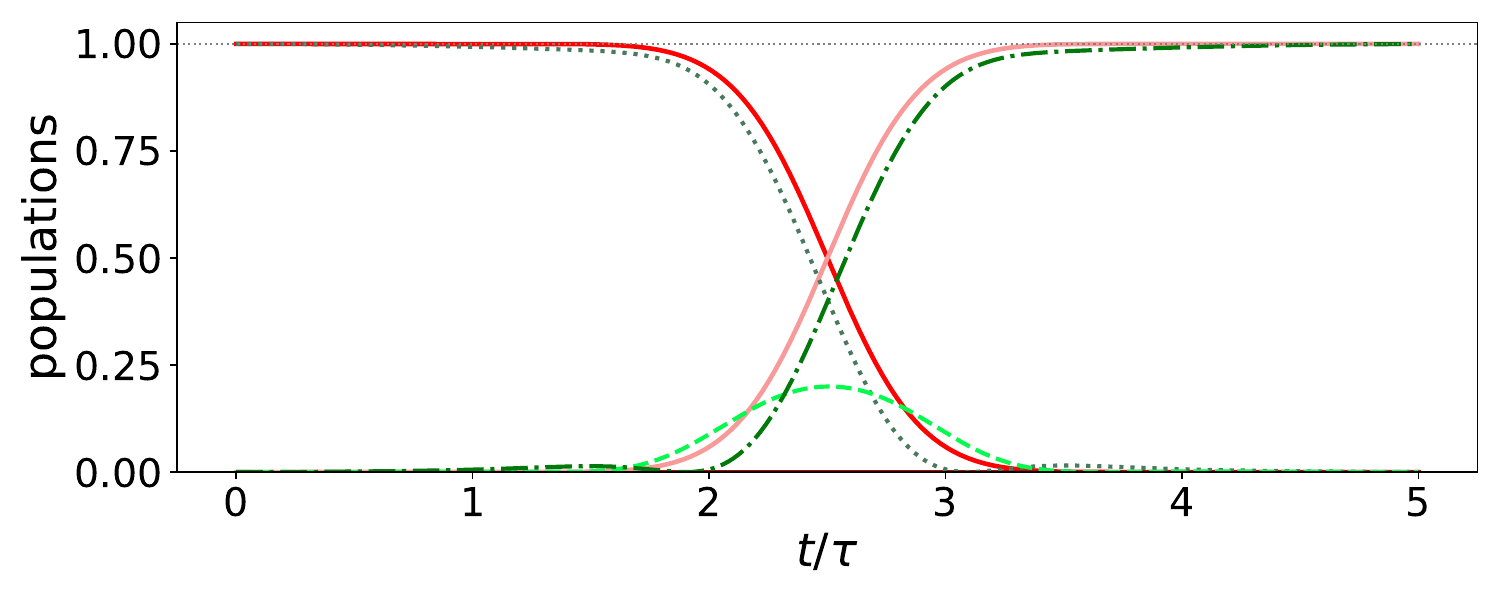}
\end{center}
\vspace{-0.7cm}
\caption{(a) Plot of the controls for the CD drive (red solid line) $v_{\text{CD}}(t) = \sqrt{2}\dot\theta(t)$, and for QOSTE$^\text{STS}$, $v_x$, $v_y$, and $v_z$ (see main text), respectively in green dots, green crosses, and green solid line. (b) Plot of the populations for the CD drive in red for $|1\ket$ and light red for $|3\ket$, and for QOSTE$^\text{STS}$, in green dots for $|1\ket$, dashed green for $|2\ket$, and dashed-dot green for $|3\ket$. %\Cami{Note that if needed one can avoid populating state $|2\ket$ by adding an extra constraint to the optimization process, as in \cite{Liu2023}}. 
Note that if needed one can additionally fix the time-integrated population in state $|2\ket$, which amounts to fixing the total loss when state $|2\ket$ is lossy, by adding an extra constraint to the optimization process, as in \cite{Liu2023}. }\label{fig:STIRAP}
\end{figure}

The CD drive associated with this problem is $V_\text{CD}(t) = i \dot \theta(t) (|1\ket\bra 3| -  |3\ket\bra 1|)$ where $\theta(t)$ is an angle related to the rotation of the instantaneous eigenvectors of $H_0(t)$ given by $\dot{\theta}=\frac{\dot{\Omega}_p\Omega_s-\dot{\Omega}_s\Omega_p}{\Omega_p^2+\Omega_s^2}$~\cite{Chen:2010}. Following the QOSTE$^\text{STS}$ method given by Eq.~\eqref{eq:vqsts}, we obtain the protocols of lowest energy cost for the STS. For $t_f = 5\tau$, $T=4\tau$, $\Omega_0=6\tau^{-1}$, and $\Delta=0$, the QOSTE$^\text{STS}$ requires only three out of the $3^2-1 =8$ possible controls, which are $v_x \sigma_x$, $v_y\sigma_y $ and $v_z\sigma_z$, with $\sigma_x = (|2\ket\bra3|+|3\ket\bra2|)\sqrt{2}$, $\sigma_y =  (i|1\ket\bra3|-i|3\ket\bra1|)\sqrt{2}$, and $\sigma_z = (|1\ket\bra2|+|2\ket\bra1|)\sqrt{2}$, and the corresponding energy cost is ${\cal C}[V_\text{Q}^{\text{STS}}] = 0.01135 ~\Omega_0 $ compared to ${\cal C}[V_\text{CD}] = 0.4725~ \Omega_0$ (leading to a ratio ${\cal C}[V_\text{CD}]/{\cal C}[V_\text{Q}^{\text{STS}}] \simeq 42$). For finite values of $\Delta$, QOSTE$^\text{STS}$ requires more controls (at least 6). We therefore propose a version of QOSTE$^\text{STS}$ with only 3 controls, $v_x$, $v_y$ and $v_z$, which is derived numerically with GRAPE.  
%Finally, we consider the same STS from $|1\ket$ to $|3\ket$ but with a reduced number of controls, $V_{\text{Q}^{STS}}=\frac{1}{2}\begin{pmatrix}
%0 & v_z & iv_y\\
%v_z & 0 & v_x\\
%-iv_y & v_x & 0
%\end{pmatrix}.$ The QOSTE$^\text{STS}$ protocol is derived numerically with a GRAPE algorithm since the analytical solution assumes full controls. 
For $\Delta = 0.1~\tau^{-1}$, the obtained QOSTE$^\text{STS}$ protocol is plotted in Fig.~\ref{fig:STIRAP}, reminding the intuitive protocol. The corresponding energy cost is ${\cal C}[V_\text{Q}^{\text{STS(3-ctrl)}}] = 0.0113777~\Omega_0$, very close to ${\cal C}[V_\text{Q}^{STS}] =0.0113769 ~\Omega_0$ with the full number of controls, while ${\cal C}[V_\text{CD}] = 0.47~ \Omega_0$ for the CD drive.\\

%\section{Conclusion}\label{secConc}
{\it \noindent Conclusion.} 
We determine, for $N$-level quantum systems, the minimum energy cost to realize arbitrary adiabatic transformations in a finite time $t_f$. The cost is equal to the length of the geodesic in the rotating frame with respect to $H_0(t)$ divided by $t_f$. We also determine explicitly the QOSTE protocol reaching this minimum energy cost, both for unitary transformation and state-to-state transfer.
%\st{For an arbitrary time-dependent Hamiltonian $H_0(t)$, we have derived by optimal control theory \st{(OCT)} and geometrical argument an energetically-optimized method, QOSTE, that realizes the adiabatic \st{evolution} \Dom{transfer} associated with $H_0(t)$, for unitary transformation as well as for state-to-state transfer.
The energy gain compared with the counterdiabatic drive grows quadratically with $t_f$.  %Analytical solutions to the optimal control problem can be derived by applying the PMP. 
%The PMP can be used to derive analytical solutions to the energy optimal control problem.
%We were able to find an analytical solution.
This is confirmed by the benchmark Landau-Zener and STIRAP models, where the energy consumed by the QOSTE is significantly smaller than the one by the CD drive, even for highly non-adiabatic situations corresponding to short $t_f$. 
Furthermore, using GRAPE we have designed a robust version of QOSTE by optimizing simultaneously the robustness and the energy of the controls; it has been shown to outperform STA, emphasizing a trade-off between robustness and energy cost.
Additionally, we highlight that the QOSTE method allows one to tailor the relative phases within the target unitary transformation, which is not possible with STA.

Moreover, the number of controls might be a critical issue for certain practical implementations. QOSTE can be derived with a reduced number of controls using GRAPE. We showed this explicitly for the STIRAP application. In the SM, we consider the case of a single control in the Landau-Zener model highlighting that although not reaching the absolute energy lower bound, it is still outperforming CD-STA.

%\st{In the Supplemental Material, we provide an example of a two-level quantum system with a constant energy gap. In this case, the difference in energy cost between QOSTE and CD drive can be very significant, reaching several orders of magnitude} \Cami{[Peut-être qu'on peut l'enlever??]}. 

%\Ste{\st{Furthermore, we compare the robustness of the QOSTE with that of the CD drive. We find that, for many but not all experimental uncertainties, the QOSTE is indeed more robust than the CD drive. Finally, we greatly improve the robustness of the QOSTE against experimental uncertainties (such as uncertainty in the control amplitude) by employing a gradient-based algorithm. Our results show that STA protocols can be made both more energy efficient and robust. They also demonstrate that increased robustness comes at an additional energetic cost.}} %In particular, we designed protocols that are much more robust than the counter-diabatic drive for the same energetic cost.

%In the Appendix, we also consider the energetically-optimized protocol with only one control. This time the optimization problem cannot be solved analytically so that we used a gradient-ascent algorithm. 

Among many perspectives, we aim to combine our findings with time-optimization and quantum speed limit procedures~\cite{delcampo21}, in particular for open systems~\cite{delCampo2013Jan,Koch2016,Alipour2020}. Our results are promising in the perspective of minimizing the energy consumption of quantum technologies~\cite{Auffeves22,Fellous-Asiani23}. \\
%\Ste{\st{Furthermore, we intend to conduct a more in-depth analysis of the emerging trade-off between robustness and energy cost.}} \\

\noindent\textbf{Acknowledgments} \\
C.L.L. acknowledges funding from the French National
Research Agency (ANR) under grant ANR-23-CPJ1-
0030-01.
The research work of D. Sugny has been supported by the ANR project ``QuCoBEC'' ANR-22-CE47-0008-02 and by the MITI CNRS project CONV. The authors acknowledge support from the EUR-EIPHI Graduate School (17-EURE-0002, SQC and CCNV projects), the QuanTEdu-France project (ANR-22-CMAS-0001), the Conseil Régional de Bourgogne-Franche-Comté, and Dijon Métropole.\\

\appendix

\vspace{1cm}
\section*{Supplemental Material}

In the Supplemental Material, we provide technical proofs of some results mentioned in the main text of the paper. In Sec.~\ref{sec:geomsol}, we derive the QOSTE protocol for unitary transformation. The different geometric arguments are described in detail. The inequalities between the energy costs, as well as the scaling of the ratio between ${\cal C}[V_\text{CD}]$ and ${\cal C}[V_\text{Q}]$ when the control time becomes very large, are demonstrated. Section~\ref{QPMP} focuses on the same control problem, but uses the tools of optimal control theory. The case of state-to-state transfer is investigated in Sec.~\ref{QSTS}.
Section~\ref{app:LZnumerics} provides numerical simulations for the study of the robustness of the counter-diabatic drive and of the QOSTE in the case of the Landau-Zener model. Section~\ref{app:singlec} presents the QOSTE for the Landau-Zener model when we assume that only a single control (along $\sigma_y$) is available, and compare it with the counterdiabatic drive. Finally, in Sec.~\ref{app:grape}, we briefly present the main lines of the used GRAPE algorithm.

\section{QOSTE for unitary transformation}\label{sec:geomsol}
%\Cami{Nouvelle section}\\

For unitary transformations, one can define the length of the trajectory described by $U(t)$ from 0 to $t$ as
\bea\label{eq:lengthU}
L_{0,t}[U] = \int_0^t du || \dot U(t)|| &=& \int_0^t du \sqrt{{\rm Tr}[\dot U^\dag(t) \dot U(t)]}\nonumber \\
&=&\int_0^t du \sqrt{{\rm Tr}[H(u)^2]},
\eea
where $H(t)$ is the Hamiltonian generating $U(t)$. Then, as shown in \cite{Wang15}, the geodesic from $\mathbb{I}$ to $U_f$ is given by $U_\text{geo}(t) = e^{-itV}$, $t\in[0;1]$, with $V = i \ln U_f$. Consequently, the length of the geodesic from $\mathbb{I}$ to $U_f$ is
\be\label{eq:length}
d(\mathbb{I}, U_f) = \sqrt{{\rm Tr}[V^2]},
\ee
with $e^{-iV} = U_f$.

\subsection{Control of constant norm}
In this section, we show that there exists a control of constant norm realizing $U(t_f) = U_f$.
We are looking for $V(t)$ such that 
\be
U(t_f) = {\cal T}e^{-i\int_0^{t_f} dt H_0(t) +V(t)} = U_f
\ee
or equivalently,
\be
\tilde U(t_f) =  {\cal T}e^{-i\int_0^{t_f} dt \tilde V(t)} = \tilde U_f := U_0^\dag(t_f) U_f
\ee
with $\tilde V(t) := U_0^\dag(t) V(t) U_0(t)$. We will then derive a protocol following \cite{Latune21}. Let us assume that $\tilde V(t) = \dot f(t) \chi$, where $\chi$ is a constant hermitian operator and $f$ a real function to be specified later on. We have
\be
\tilde U(t_f) = e^{-i [f(t_f)-f(0)] \chi}.
\ee
Then, we just need to choose $f(t_f)- f(0) =1$ and $\chi$ such that $e^{-i \chi} = \tilde U_f$. In the basis $\{\sigma_k\}$, $V(t) = U_0(t) \tilde V(t) U_0^\dag(t)$ can be spanned as
\be
V(t) = \sum_k  {\rm Tr}[\sigma_k V(t)] \sigma_k,
\ee
leading to $v_k(t) =  {\rm Tr}[\sigma_k V(t)]$, and
\bea
\sum_k v_k^2(t) &=& {\rm Tr}[V^\dag(t) V(t)]  = {\rm Tr}[\tilde V^\dag(t) \tilde V(t)] \nn\\
&=&\dot f^2(t) {\rm Tr}[ \chi^2].
\eea
Finally, choosing $f(t) = t/t_f$ (it is not the only choice), we obtain the operator 
\be\label{eq:defV}
V(t) = \frac{1}{t_f}U_0(t)\chi U_0^\dag(t)
\ee
of constant norm and fulfilling $U(t_f) = U_f$. Such operator therefore saturates the energy lower bound ${\cal C}[V] = t_f^{-1} L_{0,t_f}^2[\tilde U]$.

\subsection{Length of geodesic}
The length of the geodesic between $\tilde U(0) = \mathbb{I} $ and $\tilde U_f$ is, according to Eq. \eqref{eq:length}, 
\be
 G_{0,t_f} = \sqrt{{\rm Tr}[\chi^2]},
\ee
since $\tilde U_f = e^{-i\chi}$. Now, we compute the length of the trajectory of $\tilde U(t)$.
\bea L_{0,t_f}[\tilde U] &=& \int_0^{t_f} dt \sqrt{{\rm Tr}[\dot {\tilde U}^\dag(t) \dot {\tilde U}(t)] }\nn\\
&=&  \int_0^{t_f} dt \sqrt{{\rm Tr}[\dot {\tilde V}^\dag(t) \dot {\tilde V}(t)] }\nn\\
&=&  \int_0^{t_f} dt \frac{1}{t_f} \sqrt{ {\rm Tr}[\chi^2]}\nn\\
&=& \sqrt{{\rm Tr}[\chi^2]} \nn\\
&=&  G_{0,t_f}.
\eea
We conclude that the control introduced in Eq. \eqref{eq:defV}  follows the geodesic in the rotating frame from $\mathbb{I}$ to $\tilde U_f$. It is therefore the protocol of minimal energy cost, called the QOSTE, denoted by $V_Q$ in the main text.

\subsection{Limit when $t_f \rightarrow \infty$}

\subsubsection{Limit of $ G_{0,t_f}$ when $t_f \rightarrow \infty$}
From the previous section, we have $ G_{0,t_f} = \sqrt{{\rm Tr}[\chi^2]}$, with $ e^{-i\chi} = \tilde U_f = U_0^\dag(t_f)U_f$. It was shown in \cite{Albash12} that for $t_f \rightarrow \infty$, 
\be
U_0(t_f) = [\mathbb{I} + V_1(t_f) + {\cal O}(t_f^2)]U_\text{ad}(t_f)
\ee
where 
\be
U_\text{ad}(t) = \sum_k e^{-i \phi_k(t)}|k(t)\ket\bra k(0)|,
\ee
is precisely the adiabatic evolution, with $\phi_k(t) = \int_0^t du [e_k(u) + i\bra \dot k(u)|k(u)\ket]$ is the sum of the dynamical phase and the Berry phase. For completeness, we mention that the first order correction in $1/t_f$ is 
\bea
V_1(t) &=& - \sum_{k \ne k'} |k(0)\ket \bra k'(0)|e^{-i[\phi_{k'}(t)-\phi_{k}(t)]}\nn\\
&&\times \int_0^tdu e^{i[\phi_{k'}(u) - \phi_k(u)]}\bra k(u)|\dot k'(u)\ket.
\eea
It is shown explicitly in~\cite{Albash12} (this result is not trivial) that $V_1(t_f)$ is indeed of order $1/t_f$. Then, remembering that the target unitary $U_f$ is the adiabatic transformation, we have 
\be
 U_0^\dag(t_f)U_f = \mathbb{I} + U_\text{ad}^\dag(t_f) V_1(t_f) U_\text{ad}(t_f) + {\cal O}(t_f^{-2}).
 \ee
 This implies that, to first order (one can verify that $V_1(t)$ is indeed anti-Hermitian, $V_1^\dag(t) = - V_1(t)$),
 \be 
 -i\chi =  U_\text{ad}^\dag(t_f) V_1(t_f) U_\text{ad}(t_f) + {\cal O}(t_f^{-2}).
 \ee 
 Finally, we obtain
 \bea
  G_{0,t_f} &=& \sqrt{{\rm Tr}[\chi^2]} \nn\\
 &=& \sqrt{{\rm Tr}\{-[U_\text{ad}^\dag(t_f) V_1(t_f) U_\text{ad}(t_f)]^2\}} + {\cal O}(t_f^{-2}) \nn\\
 &=& \sqrt{{\rm Tr}\{- [V_1(t_f)]^2\}}  + {\cal O}(t_f^{-2})\nn\\
 &=& {\cal O}(t_f^{-1}).
\eea
We conclude that $ G_{t_f,0}$ scales as $t_f^{-1}$ in the adiabatic limit (when $t_f \rightarrow \infty$).

\subsubsection{Limit of $ L_{0,t_f}[\tilde U_\text{CD}]$ when $t_f \rightarrow \infty$}
We recall that  
\begin{eqnarray*}
 L_{0,t_f}[\tilde U_\text{CD}]&=&\int_0^{t_f} dt \sqrt{{\rm Tr}[\dot{\tilde U}_\text{CD}^\dag(t)\dot{\tilde U}_\text{CD}(t)]} \\
&  =& \int_0^{t_f} dt \sqrt{{\rm Tr}[V_\text{CD}^2(t)]},
\end{eqnarray*}
and that the counter-diabatic drive is given by $V_\text{CD}(t) =i \sum_n \big[|\dot n(t)\ket \bra n(t)| - \bra n(t)|\dot n(t)\ket |n(t)\ket\bra n(t)|\big]$. Then, one obtains
\be
{\rm Tr}[V_\text{CD}^2(t)] = \sum_n \Big(\bra \dot n(t)|\dot n(t)\ket - |\bra n(t)|\dot n(t)\ket|^2\Big).
\ee
We now define the unitary evolution
\be
W(t) := \sum_n e^{-\int_0^t du \bra n(u)|\dot n(u)\ket} |n(t)\ket \bra n(0)|.
\ee
One can verify that ${\rm Tr}[\dot W^\dag(t) \dot W(t)] = \sum_n \Big(\bra \dot n(t)|\dot n(t)\ket - |\bra n(t)|\dot n(t)\ket|^2\Big)$, and conclude that
\be
L_{0,t_f}[W] =  L_{0,t_f}[\tilde U_\text{CD}].
\ee
A natural lower bound of $L_{0,t_f}[W]$ is the length of the geodesic between $W(0) = \mathbb{I}$ and $W(t_f) = W_f:= \sum_n e^{-i\alpha_n^f} |n_f\ket \bra n(0)|$, where $\{|n_f\ket \}$ are the eigenstates of $H_f$, and $\alpha_n^f := \int_0^{t_f}d u\bra n(u)|\dot n(u)\ket$,
\be
L_{0,t_f}[W] \geq d(\mathbb{I},W_f) = \sqrt{{\rm Tr}[\xi^2]}
\ee
where $\xi := -i \ln[W_f]$. While $|n_f\ket$ is independent of $t_f$ because defined directly from $H_f$, the phases $\alpha_n^f$ do depend on $t_f$. Still, $d(\mathbb{I},W_f)$ is lower bounded by $d(\mathbb{I},W_\text{min})$, where $W_\text{min}$ is the unitary of the form $\sum_n e^{-i\beta_n} |n_f\ket \bra n(0)|$ closest to identity.

Then, since $d(\mathbb{I},W_\text{min}) >0$ ($W_\text{min} \ne \mathbb{I}$ since $H_f \ne H_0(0)$), we conclude that $ L_{0,t_f}[\tilde U_\text{CD}]$ is lower bounded by a constant independent of $t_f$ strictly larger than 0.

\subsubsection{${\cal C}[V_\text{CD}]$ Versus ${\cal C}[V_\text{Q}]$}
As shown in the main text, the energy cost of the STA protocol is lower bounded by the length of the trajectory in the interaction picture, $t_f{\cal C}[V_\text{CD}] \geq   L_{0,t_f}^2[\tilde U_\text{CD}]=\Big(\int_0^{t_f} dt \sqrt{{\rm Tr}[\dot{\tilde U}_\text{CD}^\dag(t)\dot{\tilde U}_\text{CD}(t)]}\Big)^2=\Big(\int_0^{t_f} dt \sqrt{{\rm Tr}[V_\text{CD}^2(t)]}\Big)^2$. On the other hand, we show that ${\cal C}[V_\text{Q}] = t_f  G_{0,t_f}^2$, leading to
\be
\frac{{\cal C}[V_\text{CD}]}{{\cal C}[V_\text{Q}]} \geq \frac{ L_{0,t_f}^2[\tilde U_\text{CD}]}{ G_{0,t_f}^2} \geq \frac{d^2(\mathbb{I}, W_\text{min})}{ G_{0,t_f}^2},
\ee
so that combining the results of the two above subsections,
\be
\frac{{\cal C}[V_\text{CD}]}{{\cal C}[V_\text{Q}]} \underset{t_f \rightarrow \infty}{\sim} A t_f^2
\ee
where $A$ is a constant independent of $t_f$.

\section{QOSTE from optimal control theory}\label{QPMP}
%\Cami{[Nouvelle section]}\\

In this section we derive the QOSTE protocol for unitary tranformation using tools of optimal control theory. The Pontryagin Maximal Principle (PMP) is derived from the minimization of an action~\cite{Boscain21,Ansel24}. Here we extend such action for optimization of unitary transformations as,
\be
S = \int_0^{t_f} dt\Big\{ \sum_k v_k^2(t) + \Re{\rm Tr}\{\Lambda^\dag(t)[\dot U(t) + i H(t)U(t)] \}\Big\},
\ee
where the $v_k(t)$ represent the controls, $H(t) = H_0(t) + \sum_k v_k(t)\sigma_k$ is the total Hamiltonian containing the original protocol $H_0(t)$, $\Lambda(t)$ is the co-unitary (equivalent to the co-state for state-to-state transfer, playing the role of Lagrange multipliers) and $U(t)$ is the unitary transformation generated by $H(t)$ to optimize. Note that we assume that the operators $\sigma_k$ form an orthonormal basis of the space of traceless operators. Note also that here we choose not to include a final cost term related to the target transformation $U_f$. Instead, we will later on determine initial condition such that $U(t_f)$ reaches $U_f$ exactly. Computing the variation $\delta S$ under variations of the $v_k(t)$, $U(t)$, $\Lambda^\dag(t)$, and $\dot U(t)$ (followed by integration by part), the condition $\delta S=0$ generates the following equations,
\bea
\dot U(t) &=& -i H(t) U(t),\nn\\
\dot \Lambda(t) &=& -i H(t)\Lambda(t),\nn\\
v_k(t) &=& \frac{1}{2}\Im {\rm Tr}\left[\Lambda^\dag(t) \sigma_k U(t)\right].
\eea
This set of equations can be re-expressed in the rotating frame with respect to $H_0(t)$, 
\bea
\dot {\tilde U}(t) &=& -i \tilde V(t) \tilde U(t),\label{eq:pmpu}\\
\dot {\tilde\Lambda}(t) &=& -i\tilde V(t) \tilde \Lambda(t),\label{eq:pmplambda}\\
v_k(t) &=& \frac{1}{2}\Im {\rm Tr}\left[\tilde \Lambda^\dag(t) \tilde \sigma_k(t) \tilde U(t)\right],\label{eq:pmpunit}
\eea
with $\tilde U(t) = U_0^\dag(t)U(t)$, $\tilde \Lambda(t) = U_0^\dag(t)\Lambda(t)$, $\tilde V(t) = U_0^\dag(t) V(t) U_0(t)$, and $\tilde \sigma(t)= U_0^\dag(t)\sigma_k U_0(t)$. Then, substituting Eq.~\eqref{eq:pmpunit} into $\tilde V(t) = \sum_k v_k(t) \tilde \sigma_k(t)$, we obtain,
\bea
\tilde V(t) &=& \frac{1}{4i}\sum_k \Big[{\rm Tr}[\tilde \Lambda(t)\tilde\sigma_k(t)\tilde U(t)] - {\rm h.c}\Big]\tilde\sigma_k(t)\nn\\
&=& \frac{1}{4i} \Big[\tilde U(t)\tilde\Lambda^\dag(t) - \tilde\Lambda(t)\tilde U^\dag(t)\Big].
\eea
Then, using Eqs.~\eqref{eq:pmplambda} and \eqref{eq:pmpu}, yields,
\bea
\dot {\tilde U}(t) &=& \frac{1}{4} \tilde\Lambda(t) -\frac{1}{4}\tilde U(t)A,\nn\\
\dot {\tilde\Lambda}(t) &=& -\frac{1}{4}\tilde U(t) +\frac{1}{4} \tilde \Lambda(t) A^\dag,\nn
\eea
with $A:=\Lambda^\dag(0)U(0)$. Using such equation, one can verify that
\be
\dot{\tilde V}(t) = 0.\nn
\ee
This implies that $\tilde V(t) = \tilde V$ is constant, so that $\tilde U(t) = e^{-i\tilde V t}$. We conclude that in order to reach the target transformation $\tilde U_f = U_0^\dag(t_f) U_f$, we must choose $\tilde V = \chi = i t_f^{-1} \ln \tilde U_f$.
We therefore recover the same solution derived from geometric arguments in Sec.~\ref{sec:geomsol}, namely the protocol of minimal energy is $V(t) = i t_f^{-1} U_0(t) \ln(\tilde U_f) U_0^\dag(t).$

\section{QOSTE for state-to-state transfer}\label{QSTS}
%\Cami{[Nouvelle section]}\\

In this section we derive the explicit expression of the least energetically expensive control for state-to-state transfer. The Hamiltonian $H_0(t)$ is given, and we denote by $H_0(0)=H_i$ and $H_0(t_f)=H_f$ the initial and final Hamiltonians. Since the CD drive is traceless, we also assume that the QOSTE protocol $V_{\text{Q}}^\text{STS}(t)$ is traceless, $\mathrm{Tr}\big(V_\text{Q}^\text{STS}(t)\big)=0$. Then, the QOSTE protocol can be decomposed as
\[
V_\text{Q}^\text{STS}(t)=\sum_k v_k(t)\sigma_k
\]
where $\sigma_k$ is a basis of traceless Hermitian operators with $\mathrm{Tr}(\sigma_k\sigma_{k'})=\delta_{k,k'}$. The energy cost to minimize is
\[
\mathcal{C}[V_\text{Q}^\text{STS}]
=
\int_0^{t_f} dt \;
\mathrm{Tr}\left[\left(V_\text{Q}^\text{STS}(t)\right)^2\right]
=
\sum_k
\int_0^{t_f} dt \; v_k^2(t).
\]
The PMP gives \cite{pontryaginbook,leemarkusbook,Ansel24,Boscain21,bonnard_optimal_2012}
\begin{align}
|\dot{\psi}(t)\ket &= -i H(t)|\psi(t)\ket,\nn \\
|\dot{\chi}(t)\ket &= -i H(t)|\chi(t)\ket,\nn \\
2v_k(t) &= \Im\,\mathrm{Tr}
\left(
\bra\chi(t)|\sigma_k|\psi(t)\ket
\right).\nn
\end{align}
Note that we did not include a terminal cost to enforce
$|\psi(t_f)\ket=|\psi_f\ket$. Instead, we will determine initial conditions such that the target is reached exactly. Since
\[
H(t)=H_0(t)+V_\text{Q}^\text{STS}(t),
\]
we move to the rotating frame with respect to $H_0(t)$:
\begin{align}
|\tilde{\psi}(t)\ket &= U_0^\dagger(t)|\psi(t)\ket,\nn\\
|\tilde{\chi}(t)\ket &= U_0^\dagger(t)|\chi(t)\ket,\nn\\
|\tilde{\psi}_f\ket &= U_0^\dagger(t_f)|\psi_f\ket.\nn
\end{align}
Then
\begin{align}
|\dot{\tilde{\psi}}(t)\ket &= -i \tilde V_\text{Q}^\text{STS}(t)|\tilde{\psi}(t)\ket,\nn\\
|\dot{\tilde{\chi}}(t)\ket &= -i \tilde V_\text{Q}^\text{STS}(t)|\tilde{\chi}(t)\ket .\nn
\end{align}
and
\[
v_k(t)=\frac{1}{2}
\Im\,
\mathrm{Tr}
\left(
\bra\tilde{\chi}(t)|\tilde{\sigma}_k(t)|\tilde{\psi}(t)\ket
\right),
\]
with $\tilde{\sigma}_k(t)=U_0^\dagger(t)\sigma_k U_0(t)$. On the one hand, we have 
\begin{align}
|\tilde{\psi}(t)\ket &= \tilde U(t)|\psi(0)\ket,\nn\\
|\tilde{\chi}(t)\ket &= \tilde U(t)|\chi(0)\ket,\nn
\end{align}
with
\[
\tilde U(t)
=
\mathcal{T}
\exp
\left(
-i
\int_0^t du \;
\tilde V_Q(u)
\right).
\]
On the other hand, since $V_\text{Q}^\text{STS}(t)=\sum_k v_k(t)\sigma_k$, we obtain 
\[
\tilde V_\text{Q}^\text{STS}(t)=U_0^\dagger(t)V_\text{Q}^\text{STS}(t)U_0(t)
=
\sum_k v_k(t)\tilde\sigma_k(t).
\]
Hence
\begin{align}
\tilde V_\text{Q}^\text{STS}(t)
&=
\sum_k
\frac{1}{2}
\Im
\mathrm{Tr}
\big(
\bra\tilde{\chi}(t) |\tilde{\sigma}_k(t) |\tilde{\psi}(t)\ket
\big)
\tilde{\sigma}_k(t)\nn
\\
&=
\frac{1}{4i}
\sum_k
\left[
\mathrm{Tr}
\big(
\bra \tilde{\chi}(t)|
\tilde{\sigma}_k(t)
|\tilde{\psi}(t)\ket
\big)
\tilde{\sigma}_k(t)
- h.c.
\right].\nn\\
&=\frac{1}{4i}
\sum_k
\left[
\mathrm{Tr}
\big( \tilde{\sigma}_k(t)
|\tilde{\psi}(t)\ket\bra \tilde{\chi}(t)|
\big)
\tilde{\sigma}_k(t)
- h.c.
\right].\nn\\
&=\frac{1}{4i}
\left[
|\tilde{\psi}(t)\ket\bra\tilde{\chi}(t)|
-
|\tilde{\chi}(t)\ket\bra\tilde{\psi}(t)|
\right],
\end{align}
since $V_\text{Q}^\text{STS}(t)$ is traceless. We define
\be
|\tilde{\phi}(t)\ket
=
\frac{
|\tilde{\chi}(t)\ket
-
\bra\tilde{\psi}(t)|\tilde{\chi}(t)\ket
|\tilde{\psi}(t)\ket
}{
\sqrt{
\bra\tilde{\chi}(t)|\tilde{\chi}(t)\ket
-
|\bra\tilde{\psi}(t)|\tilde{\chi}(t)\ket|^2}},
\ee
that implies $\bra\tilde{\phi}(t)|\tilde{\psi}(t)\ket=0$. Let $\bra\tilde{\chi}(t)|\tilde{\chi}(t)\ket = \bra\chi(0)|\chi(0)\ket = c$, and $\bra\tilde{\psi}(t)|\tilde{\chi}(t)\ket = \bra\psi(0)|\chi(0)\ket = d$. Hence we have $|\tilde{\chi}(t)\ket = R|\tilde{\phi}(t)\ket + d|\tilde{\psi}(t)\ket$, with $R=\sqrt{c-|d|^2}$. Substituting into $\tilde V_\text{Q}^\text{STS}(t)$ gives $\tilde V_\text{Q}^\text{STS}(t) = \frac{R}{4i}\Big(|\tilde \psi(t)\ket\bra \tilde \phi(t)|  -|\tilde \phi(t)\ket\bra \tilde \psi(t)|\Big)- \frac{1}{2}\Im (d) |\tilde \psi(t)\ket\bra \tilde \psi(t)|$.
Therefore we get
\bea
|\dot{\tilde{\psi}}(t)\ket &=& -i\tilde V_\text{Q}^\text{STS}(t) |\tilde \psi(t)\ket \nn\\
&=& \frac{R}{4}|\tilde \phi(t)\ket + i \frac{\Im(d)}{2}|\tilde \psi(t)]\ket,\nn
\eea
and
\bea
|\dot{\tilde{\phi}}(t)\ket&=& -i\tilde V_\text{Q}^\text{STS}(t) |\tilde \phi(t)\ket \nn\\
&=& -\frac{R}{4}|\tilde{\psi}(t)\ket.\nn
\eea
This can be written as
\be
\frac{d}{dt}
\begin{pmatrix}
|\tilde{\psi}(t)\ket\\
|\tilde{\phi}(t)\ket
\end{pmatrix}
=
A
\begin{pmatrix}
|\tilde{\psi}(t)\ket\\
|\tilde{\phi}(t)\ket
\end{pmatrix},\nn
\ee
with
\be
A=
\begin{pmatrix}
i\frac{\Im(d)}{2} & -\frac{R}{4}\\
\frac{R}{4} & 0
\end{pmatrix}.\nn
\ee
The eigenvalues follow from $|A-\lambda I|=0$, which gives
\be
\lambda_\pm
=
\frac{i\,\Im(d)}{4}
\pm
\frac{i}{4}
\sqrt{\Im(d)^2+R^2}.\nn
\ee
Note that for any state $\rho_\text{diag}$ diagonal in the eigen-energy basis of $H_0(t)$ at the instant $t$, ${\rm Tr}[\rho_\text{diag} V_\text{CD}(t)]=0$: the counter-diabatic drive does not affect the local energy of the system. One can impose the same property for the QOSTE protocol. Assuming $|\psi(0)\ket$ is one of the eigenstate of $H_0(0)$, ${\rm Tr}[\rho_\text{diag} V_\text{Q}^\text{STS}(t)] = 0$ is guaranteed simply by choosing $\Im (d) = 0$.

With this condition, $\lambda_\pm$ becomes $\lambda_\pm = \pm i R/4$, and the solution of the above system is simply
\bea
|\tilde \psi(t)\ket &=& \cos\left(\frac{Rt}{4}\right)|\psi(0)\ket + \sin\left(\frac{Rt}{4}\right)|\phi(0)\ket,\nn\\
|\tilde \phi(t)\ket &=&-\sin\left(\frac{Rt}{4}\right)|\psi(0)\ket + \cos\left(\frac{Rt}{4}\right)|\phi(0)\ket,\nn
\eea
and 
\be\label{eqsm:vinterm}
\tilde V_\text{Q}^\text{STS}(t) = -i\frac{R}{4}\left( |\psi(0)\ket\bra \phi(0)| - |\phi(0)\ket\bra \psi(0)|\right),
\ee
which is in fact time-independent. We now have to choose $R$ and $|\phi(0)\ket$ such that $|\tilde \psi(t_f\ket = |\tilde \psi_f\ket$. Let us introduce
\be
|\psi_\perp\ket = \frac{(e^{-i\varphi_f}|\tilde \psi_f\ket - |\bra \psi(0)|\tilde \psi_f\ket||\psi(0)\ket)}{\sqrt{1-|\bra \psi(0)|\tilde \psi_f\ket|^2}},\nn
\ee
which is orthonormal to $|\psi(0)\ket$, and where $\varphi_f = \arg\bra\psi(0)|\tilde\psi_f\ket $. Introducing $\alpha_f:= \arccos|\bra\psi(0)|\tilde\psi_f\ket|$, we have 
\be
|\tilde \psi_f\ket = e^{i\varphi_f} \left(\sin\alpha_f |\psi_\perp\ket + \cos\alpha_f|\psi(0)\ket\right).\nn
\ee
Then, choosing $|\phi(0)\ket = |\psi_\perp\ket$ and $R = 4\alpha_f/t_f$ and substituting in Eq. \eqref{eqsm:vinterm} we can verify that \be
\tilde V_\text{Q}^\text{STS}(t) = -i\frac{\alpha_f}{t_f}\left( |\psi(0)\ket\bra \psi_\perp| - |\psi_\perp\ket\bra \psi(0)|\right),\nn
\ee
indeed takes $|\psi(0)\ket$ to $e^{-i\varphi_f}|\tilde \psi_f\ket$. Computing the energy cost of such control, one obtains
\bea
{\cal C}[V_\text{Q}^\text{STS}] &=& \int_0^{t_f}dt {\rm Tr}\left[\left(V_\text{Q}^\text{STS}(t)\right)^2\right] = \int_0^{t_f}dt {\rm Tr}\left[\left(\tilde V_\text{Q}^\text{STS}\right)^2\right]\nn\\
&=& 2 \alpha_f^2/t_f.\nn
\eea
Note that since $\alpha_f= \arccos|\bra\psi(0)|\tilde\psi_f\ket|$, which is equal to $G_{0,t_f}^\text{STS}$, the length of the geodesic between $|\psi(0)\ket$ and $|\tilde \psi_f\ket$, the energy cost of $V_\text{Q}^\text{STS}$ can also be expressed as ${\cal C}[V_\text{Q}^\text{STS}] 
= 2 [G_{0,t_f}^\text{STS}]^2/t_f$.

We observe that with the above protocol $\tilde V_\text{Q}^\text{STS}$, one reaches $|\tilde \psi_f\ket$ up to the global phase $\varphi_f$. If one needs to reach $|\tilde \psi_f\ket$ including the global phase, one can add the term $-\varphi_f/t_f \mathbb{I}$ to $\tilde V_\text{Q}^\text{STS}$. Note that this will increase the energy cost. Alternatively, one can use the QOSTE protocol for unitary transformations where all relative phases can be controlled at will.

\section{Numerical results on robustness for the Landau-Zener model}\label{app:LZnumerics}
\begin{figure}
\begin{center}
   (a) \includegraphics[width=0.45\textwidth]{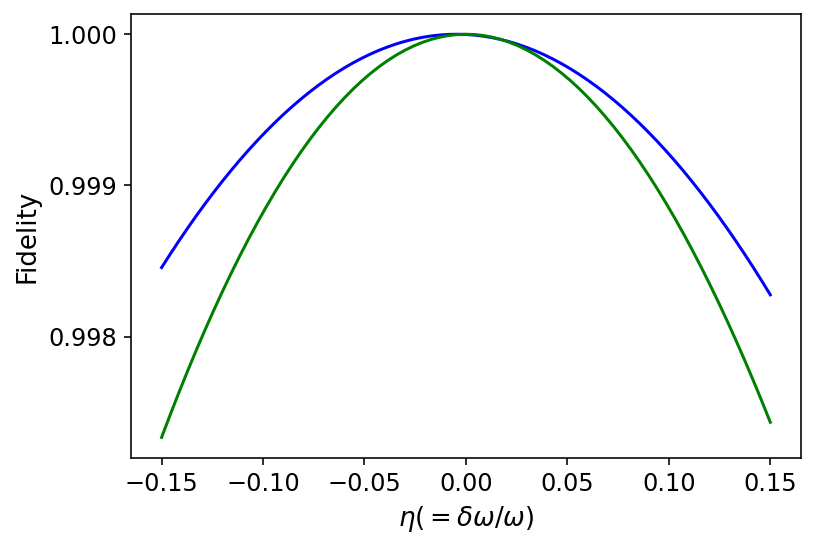}  \\
(b)  \includegraphics[width=0.45\textwidth]{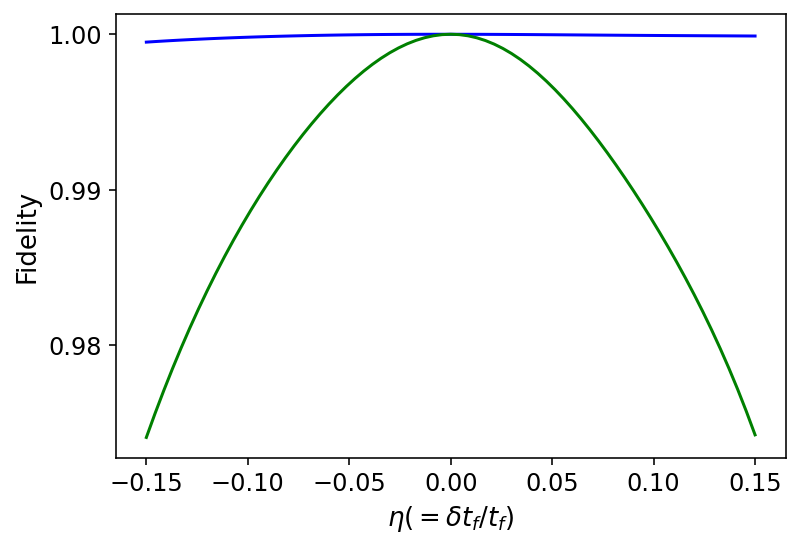}
\end{center}
\caption{Robustness of the counter-diabatic drive (in blue) compared with the robustness of the QOSTE (in green), for uncertainty (a) in the Hamiltonian parameter $\omega$; (b) in the duration of the operation $t_f$. We use $\Delta(t) = \Delta_0 +\Delta_d t/t_f$ with $\Delta_0/\omega = -10$, $\Delta_d/\omega = 20$ and $\omega t_f = 1$.}\label{fig:LZ_rob2}
\end{figure}

In this section, we present some numerical results for the robustness of the counter-diabatic drive compared to the robustness of the QOSTE for the Landau-Zener model.
We consider static errors, where the Hamiltonian parameter is not exactly known, but is in a given interval fixed by the experimental setup. 
In Fig. \ref{fig:LZ_rob2}, we show the uncertainty with respect to (a) $\omega$ and (b) the duration of the operation $t_f$, meaning that we plot the final fidelity with respect to $|\psi_\text{target}\ket$ for some altered values of $\omega \rightarrow \omega(1 + \eta) $ and $t_f \rightarrow t_f(1 + \eta) $ with $\eta \in [-0.15;0.15]$. For these two kinds of uncertainty, one can see that the counter-diabatic drive is slightly more robust than the QOSTE.

\begin{figure}
\begin{center}
   (a) \includegraphics[width=0.45\textwidth]{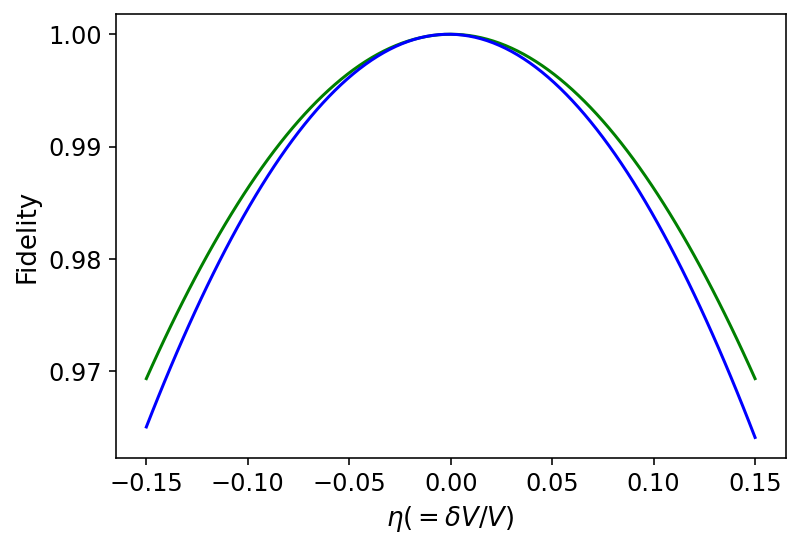}  \\
(b)  \includegraphics[width=0.45\textwidth]{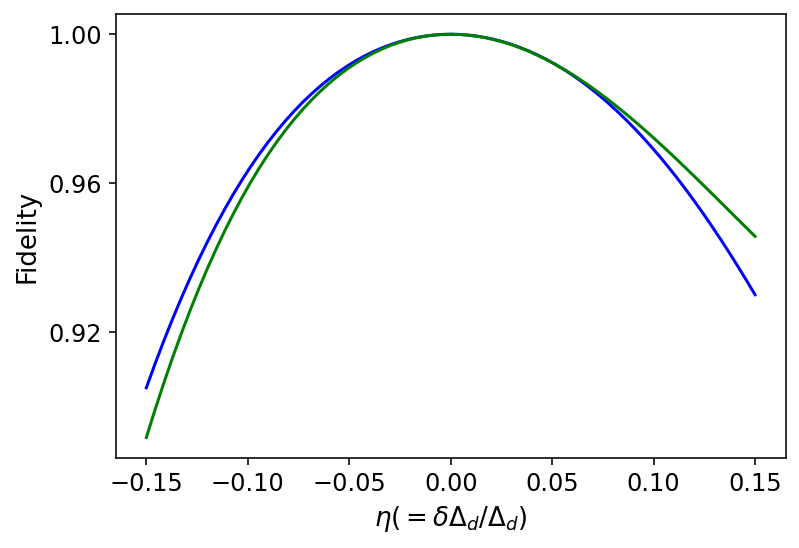}
\end{center}
\caption{Robustness of the counter-diabatic drive (in blue) compared with the robustness of the QOSTE (in green), for uncertainty (a) in the amplitude of the control; (b) in the amplitude of $\Delta_d$. % $H_0(t)$.
We used $\Delta(t) = \Delta_0 +\Delta_d t/t_f$ with $\Delta_0/\omega = -10$, $\Delta_d/\omega = 20$ and $\omega t_f = 1$. }\label{fig:LZ_rob}
\end{figure}

In Fig.~\ref{fig:LZ_rob}(a), we plot the robustness with respect to uncertainty in the amplitude of the controls, meaning that we plot the final fidelity with respect to $|\psi_\text{target}\ket$ for some uncertainties $\eta$ with respect to the amplitude of the control functions, $(1+\eta)V_\text{Q}^\text{STS}(t)$ instead of $V_\text{Q}^\text{STS}(t)$, and $(1+\eta)V_{CD}(t)$ instead of $(1+\eta)V_{CD}(t)$ for the counter-diabatic drive.
 In Fig.~\ref{fig:LZ_rob}(b), we represent the robustness of the protocols with respect to uncertainty in the amplitude of the parameter $\Delta_d$ (which implies that the counter-diabatic drive and the QOSTE are determined with a value of $\Delta_d$ which is not the exact one). We observe that for these uncertainties, the QOSTE is slightly more robust than the counter-diabatic drive.

\section{A case study of a single control in the Landau-Zener model}\label{app:singlec}
\begin{figure}
\begin{center}
  \includegraphics[width=0.45\textwidth]{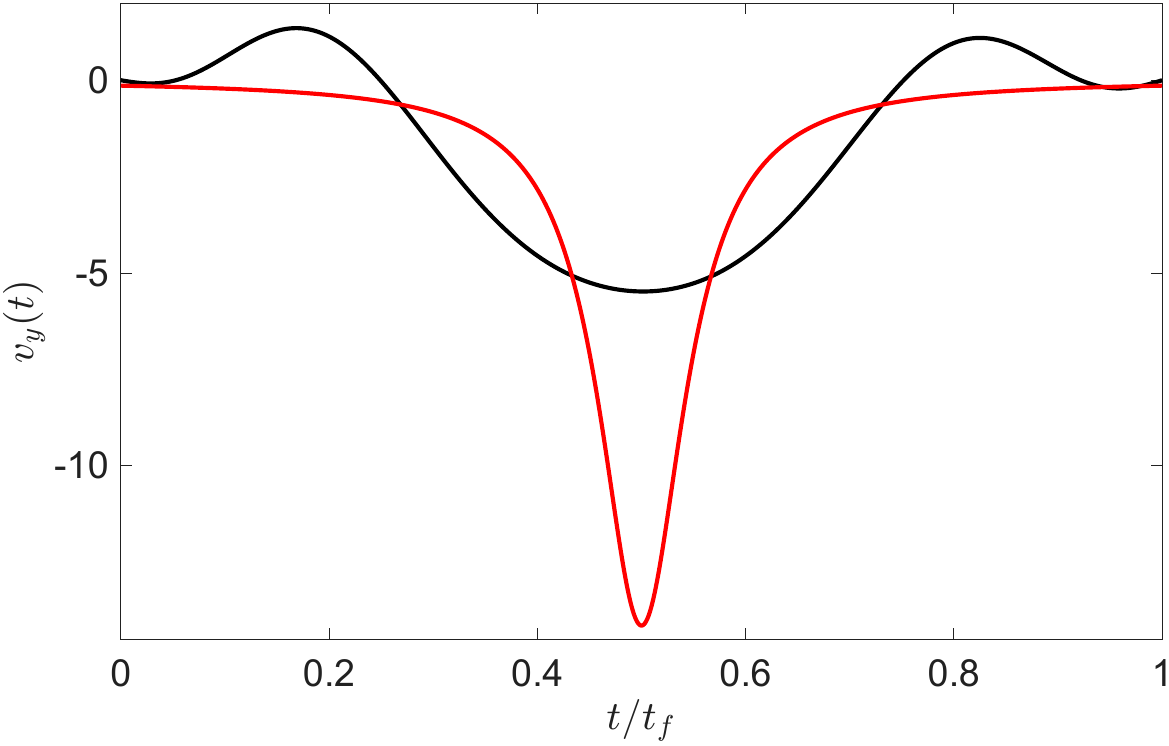}
\end{center}
\caption{Plot of the optimal control when restricted to a single control along $y$- direction (in black). For comparison, the counter-diabatic drive is also represented (in red). We used $\Delta(t) = \Delta_0 +\Delta_d t/t_f$ with $\Delta_0/\omega = -10$, $\Delta_d/\omega = 20$ and $\omega t_f = 1$.}\label{fig:single_ctrl}
\end{figure}

In this section, we study the energy optimization when only one control in the $y$-direction is optimized, rather than the three controls considered so far. In this case, the PMP equations cannot be solved analytically and a gradient-based algorithm has been used. % (see also section~\ref{app:grape}). 
The resulting optimal control is plotted in Fig.~\ref{fig:single_ctrl}, which corresponds to an energy cost of ${\cal C}[V_Q^\text{STS, (1-ctrl)}]=8 \omega$ (remembering that the cost for the counter-diabatic drive is equal to ${\cal C}[V_\text{CD}]= 15.7 \omega$). Note that we added the constraint of null control at the initial and final times in the numerical optimization. 

\section{Gradient-ascent-based algorithm (GRAPE)}\label{app:grape}
We present in this section a gradient-based algorithm for robust state-to-state transfer. 

Starting from the excited state $|e_i\rangle$ of the Hamiltonian $H_0(0)$, the goal of the control is to bring the system to the target excited state $|e_f\rangle$ at time $t_f$, while minimizing the cost functional $\int_0^{t_f}\vec{v}^2dt$ for any value of the unknown parameter $\eta\in [-\epsilon,\epsilon]$. The distance to the target state is given by $|\langle e_f|\psi_\eta(t_f)\rangle |^2$. We consider only $N_\eta$ values of $\eta$ in the interval $[-\epsilon,\epsilon]$ and we define the average fidelity as
$$
\bar F_\text{STS}=\frac{1}{N_\eta}\sum_\eta |\langle e_f|\psi_\eta(t_f)\rangle |^2.
$$
Since the objective of the control is to minimize both the distance to the target and the energy control, we fix the amplitude of the control to the value $v_0$, $v_0^2=\vec{v}^2(t)$.

We then apply a standard GRAPE algorithm on this system~\cite{khaneja_optimal_2005} for the simultaneous control of the $N_\eta$ kets $|\psi_\eta\rangle$~\cite{kobzar2004exploring,kobzar2012exploring}. At each iteration of the algorithm, we multiply the new control at time $t$ by the factor $\frac{u_0}{\sqrt{\vec{v}^2}}$ to ensure that the control amplitude remains constant.

\bibliography{biblio}

\begin{thebibliography}{79}%
\makeatletter
\providecommand \@ifxundefined [1]{%
 \@ifx{#1\undefined}
}%
\providecommand \@ifnum [1]{%
 \ifnum #1\expandafter \@firstoftwo
 \else \expandafter \@secondoftwo
 \fi
}%
\providecommand \@ifx [1]{%
 \ifx #1\expandafter \@firstoftwo
 \else \expandafter \@secondoftwo
 \fi
}%
\providecommand \natexlab [1]{#1}%
\providecommand \enquote  [1]{``#1''}%
\providecommand \bibnamefont  [1]{#1}%
\providecommand \bibfnamefont [1]{#1}%
\providecommand \citenamefont [1]{#1}%
\providecommand \href@noop [0]{\@secondoftwo}%
\providecommand \href [0]{\begingroup \@sanitize@url \@href}%
\providecommand \@href[1]{\@@startlink{#1}\@@href}%
\providecommand \@@href[1]{\endgroup#1\@@endlink}%
\providecommand \@sanitize@url [0]{\catcode `\\12\catcode `\$12\catcode
  `\&12\catcode `\#12\catcode `\^12\catcode `\_12\catcode `\%12\relax}%
\providecommand \@@startlink[1]{}%
\providecommand \@@endlink[0]{}%
\providecommand \url  [0]{\begingroup\@sanitize@url \@url }%
\providecommand \@url [1]{\endgroup\@href {#1}{\urlprefix }}%
\providecommand \urlprefix  [0]{URL }%
\providecommand \Eprint [0]{\href }%
\providecommand \doibase [0]{https://doi.org/}%
\providecommand \selectlanguage [0]{\@gobble}%
\providecommand \bibinfo  [0]{\@secondoftwo}%
\providecommand \bibfield  [0]{\@secondoftwo}%
\providecommand \translation [1]{[#1]}%
\providecommand \BibitemOpen [0]{}%
\providecommand \bibitemStop [0]{}%
\providecommand \bibitemNoStop [0]{.\EOS\space}%
\providecommand \EOS [0]{\spacefactor3000\relax}%
\providecommand \BibitemShut  [1]{\csname bibitem#1\endcsname}%
\let\auto@bib@innerbib\@empty
%</preamble>
\bibitem [{\citenamefont {Glaser}\ \emph {et~al.}(2015)\citenamefont {Glaser},
  \citenamefont {Boscain}, \citenamefont {Calarco}, \citenamefont {Koch},
  \citenamefont {K{\ifmmode\ddot{o}\else\"{o}\fi}ckenberger}, \citenamefont
  {Kosloff}, \citenamefont {Kuprov}, \citenamefont {Luy}, \citenamefont
  {Schirmer}, \citenamefont {Schulte-Herbr{\ifmmode\ddot{u}\else\"{u}\fi}ggen},
  \citenamefont {Sugny},\ and\ \citenamefont {Wilhelm}}]{Glaser15}%
  \BibitemOpen
  \bibfield  {author} {\bibinfo {author} {\bibfnamefont {S.~J.}\ \bibnamefont
  {Glaser}}, \bibinfo {author} {\bibfnamefont {U.}~\bibnamefont {Boscain}},
  \bibinfo {author} {\bibfnamefont {T.}~\bibnamefont {Calarco}}, \bibinfo
  {author} {\bibfnamefont {C.~P.}\ \bibnamefont {Koch}}, \bibinfo {author}
  {\bibfnamefont {W.}~\bibnamefont
  {K{\ifmmode\ddot{o}\else\"{o}\fi}ckenberger}}, \bibinfo {author}
  {\bibfnamefont {R.}~\bibnamefont {Kosloff}}, \bibinfo {author} {\bibfnamefont
  {I.}~\bibnamefont {Kuprov}}, \bibinfo {author} {\bibfnamefont
  {B.}~\bibnamefont {Luy}}, \bibinfo {author} {\bibfnamefont {S.}~\bibnamefont
  {Schirmer}}, \bibinfo {author} {\bibfnamefont {T.}~\bibnamefont
  {Schulte-Herbr{\ifmmode\ddot{u}\else\"{u}\fi}ggen}}, \bibinfo {author}
  {\bibfnamefont {D.}~\bibnamefont {Sugny}},\ and\ \bibinfo {author}
  {\bibfnamefont {F.~K.}\ \bibnamefont {Wilhelm}},\ }\bibfield  {title}
  {\bibinfo {title} {{Training Schr{\ifmmode\ddot{o}\else\"{o}\fi}dinger{'}s
  cat: quantum optimal control}},\ }\href
  {https://doi.org/10.1140/epjd/e2015-60464-1} {\bibfield  {journal} {\bibinfo
  {journal} {Eur. Phys. J. D}\ }\textbf {\bibinfo {volume} {69}},\ \bibinfo
  {pages} {1} (\bibinfo {year} {2015})}\BibitemShut {NoStop}%
\bibitem [{\citenamefont {Koch}\ \emph {et~al.}(2022)\citenamefont {Koch},
  \citenamefont {Boscain}, \citenamefont {Calarco}, \citenamefont {Dirr},
  \citenamefont {Filipp}, \citenamefont {Glaser}, \citenamefont {Kosloff},
  \citenamefont {Montangero}, \citenamefont {Schulte-Herbr\"uggen},
  \citenamefont {Sugny},\ and\ \citenamefont {Wilhelm}}]{kochroadmap}%
  \BibitemOpen
  \bibfield  {author} {\bibinfo {author} {\bibfnamefont {C.~P.}\ \bibnamefont
  {Koch}}, \bibinfo {author} {\bibfnamefont {U.}~\bibnamefont {Boscain}},
  \bibinfo {author} {\bibfnamefont {T.}~\bibnamefont {Calarco}}, \bibinfo
  {author} {\bibfnamefont {G.}~\bibnamefont {Dirr}}, \bibinfo {author}
  {\bibfnamefont {S.}~\bibnamefont {Filipp}}, \bibinfo {author} {\bibfnamefont
  {S.~J.}\ \bibnamefont {Glaser}}, \bibinfo {author} {\bibfnamefont
  {R.}~\bibnamefont {Kosloff}}, \bibinfo {author} {\bibfnamefont
  {S.}~\bibnamefont {Montangero}}, \bibinfo {author} {\bibfnamefont
  {T.}~\bibnamefont {Schulte-Herbr\"uggen}}, \bibinfo {author} {\bibfnamefont
  {D.}~\bibnamefont {Sugny}},\ and\ \bibinfo {author} {\bibfnamefont {F.~K.}\
  \bibnamefont {Wilhelm}},\ }\bibfield  {title} {\bibinfo {title} {Quantum
  optimal control in quantum technologies. strategic report on current status,
  visions and goals for research in europe},\ }\href
  {https://doi.org/10.1140/epjqt/s40507-022-00138-x} {\bibfield  {journal}
  {\bibinfo  {journal} {EPJ Quantum Technology}\ }\textbf {\bibinfo {volume}
  {9}},\ \bibinfo {pages} {19} (\bibinfo {year} {2022})}\BibitemShut {NoStop}%
\bibitem [{\citenamefont {Brif}\ \emph {et~al.}(2010)\citenamefont {Brif},
  \citenamefont {Chakrabarti},\ and\ \citenamefont
  {Rabitz}}]{past-present-future}%
  \BibitemOpen
  \bibfield  {author} {\bibinfo {author} {\bibfnamefont {C.}~\bibnamefont
  {Brif}}, \bibinfo {author} {\bibfnamefont {R.}~\bibnamefont {Chakrabarti}},\
  and\ \bibinfo {author} {\bibfnamefont {H.}~\bibnamefont {Rabitz}},\
  }\bibfield  {title} {\bibinfo {title} {Control of quantum phenomena: past,
  present and future},\ }\href@noop {} {\bibfield  {journal} {\bibinfo
  {journal} {New Journal of Physics}\ }\textbf {\bibinfo {volume} {12}},\
  \bibinfo {pages} {075008} (\bibinfo {year} {2010})}\BibitemShut {NoStop}%
\bibitem [{\citenamefont {Altafini}\ and\ \citenamefont
  {Ticozzi}(2012)}]{altafini2012}%
  \BibitemOpen
  \bibfield  {author} {\bibinfo {author} {\bibfnamefont {C.}~\bibnamefont
  {Altafini}}\ and\ \bibinfo {author} {\bibfnamefont {F.}~\bibnamefont
  {Ticozzi}},\ }\bibfield  {title} {\bibinfo {title} {Modeling and control of
  quantum systems: An introduction},\ }\href@noop {} {\bibfield  {journal}
  {\bibinfo  {journal} {IEEE Trans. Automat. Control}\ }\textbf {\bibinfo
  {volume} {57}},\ \bibinfo {pages} {1898} (\bibinfo {year}
  {2012})}\BibitemShut {NoStop}%
\bibitem [{\citenamefont {Dong}\ and\ \citenamefont
  {Petersen}(2010)}]{dong2010}%
  \BibitemOpen
  \bibfield  {author} {\bibinfo {author} {\bibfnamefont {D.}~\bibnamefont
  {Dong}}\ and\ \bibinfo {author} {\bibfnamefont {I.~A.}\ \bibnamefont
  {Petersen}},\ }\bibfield  {title} {\bibinfo {title} {Quantum control theory
  and applications: A survey},\ }\href@noop {} {\bibfield  {journal} {\bibinfo
  {journal} {IET Control Theory A}\ }\textbf {\bibinfo {volume} {4}},\ \bibinfo
  {pages} {2651} (\bibinfo {year} {2010})}\BibitemShut {NoStop}%
\bibitem [{\citenamefont {Koch}\ \emph {et~al.}(2019)\citenamefont {Koch},
  \citenamefont {Lemeshko},\ and\ \citenamefont {Sugny}}]{RMPsugny}%
  \BibitemOpen
  \bibfield  {author} {\bibinfo {author} {\bibfnamefont {C.~P.}\ \bibnamefont
  {Koch}}, \bibinfo {author} {\bibfnamefont {M.}~\bibnamefont {Lemeshko}},\
  and\ \bibinfo {author} {\bibfnamefont {D.}~\bibnamefont {Sugny}},\ }\bibfield
   {title} {\bibinfo {title} {Quantum control of molecular rotation},\ }\href
  {https://doi.org/10.1103/RevModPhys.91.035005} {\bibfield  {journal}
  {\bibinfo  {journal} {Rev. Mod. Phys.}\ }\textbf {\bibinfo {volume} {91}},\
  \bibinfo {pages} {035005} (\bibinfo {year} {2019})}\BibitemShut {NoStop}%
\bibitem [{\citenamefont {Gu{\ifmmode\acute{e}\else\'{e}\fi}ry-Odelin}\ \emph
  {et~al.}(2019)\citenamefont {Gu{\ifmmode\acute{e}\else\'{e}\fi}ry-Odelin},
  \citenamefont {Ruschhaupt}, \citenamefont {Kiely}, \citenamefont
  {Torrontegui}, \citenamefont
  {Mart{\ifmmode\acute{\imath}\else\'{\i}\fi}nez-Garaot},\ and\ \citenamefont
  {Muga}}]{Guery-Odelin19}%
  \BibitemOpen
  \bibfield  {author} {\bibinfo {author} {\bibfnamefont {D.}~\bibnamefont
  {Gu{\ifmmode\acute{e}\else\'{e}\fi}ry-Odelin}}, \bibinfo {author}
  {\bibfnamefont {A.}~\bibnamefont {Ruschhaupt}}, \bibinfo {author}
  {\bibfnamefont {A.}~\bibnamefont {Kiely}}, \bibinfo {author} {\bibfnamefont
  {E.}~\bibnamefont {Torrontegui}}, \bibinfo {author} {\bibfnamefont
  {S.}~\bibnamefont {Mart{\ifmmode\acute{\imath}\else\'{\i}\fi}nez-Garaot}},\
  and\ \bibinfo {author} {\bibfnamefont {J.~G.}\ \bibnamefont {Muga}},\
  }\bibfield  {title} {\bibinfo {title} {{Shortcuts to adiabaticity: Concepts,
  methods, and applications}},\ }\href
  {https://doi.org/10.1103/RevModPhys.91.045001} {\bibfield  {journal}
  {\bibinfo  {journal} {Rev. Mod. Phys.}\ }\textbf {\bibinfo {volume} {91}},\
  \bibinfo {pages} {045001} (\bibinfo {year} {2019})}\BibitemShut {NoStop}%
\bibitem [{\citenamefont {Stefanatos}\ and\ \citenamefont
  {Paspalakis}(2021)}]{Stefanatos2020}%
  \BibitemOpen
  \bibfield  {author} {\bibinfo {author} {\bibfnamefont {D.}~\bibnamefont
  {Stefanatos}}\ and\ \bibinfo {author} {\bibfnamefont {E.}~\bibnamefont
  {Paspalakis}},\ }\bibfield  {title} {\bibinfo {title} {A shortcut tour of
  quantum control methods for modern quantum technologies},\ }\href
  {https://doi.org/10.1209/0295-5075/132/60001} {\bibfield  {journal} {\bibinfo
   {journal} {Europhysics Letters}\ }\textbf {\bibinfo {volume} {132}},\
  \bibinfo {pages} {60001} (\bibinfo {year} {2021})}\BibitemShut {NoStop}%
\bibitem [{\citenamefont {Torrontegui}\ \emph {et~al.}(2013)\citenamefont
  {Torrontegui}, \citenamefont {Ibáñez}, \citenamefont {Martínez-Garaot},
  \citenamefont {Modugno}, \citenamefont {{del Campo}}, \citenamefont
  {Guéry-Odelin}, \citenamefont {Ruschhaupt}, \citenamefont {Chen},\ and\
  \citenamefont {Muga}}]{TORRONTEGUI2013}%
  \BibitemOpen
  \bibfield  {author} {\bibinfo {author} {\bibfnamefont {E.}~\bibnamefont
  {Torrontegui}}, \bibinfo {author} {\bibfnamefont {S.}~\bibnamefont
  {Ibáñez}}, \bibinfo {author} {\bibfnamefont {S.}~\bibnamefont
  {Martínez-Garaot}}, \bibinfo {author} {\bibfnamefont {M.}~\bibnamefont
  {Modugno}}, \bibinfo {author} {\bibfnamefont {A.}~\bibnamefont {{del
  Campo}}}, \bibinfo {author} {\bibfnamefont {D.}~\bibnamefont
  {Guéry-Odelin}}, \bibinfo {author} {\bibfnamefont {A.}~\bibnamefont
  {Ruschhaupt}}, \bibinfo {author} {\bibfnamefont {X.}~\bibnamefont {Chen}},\
  and\ \bibinfo {author} {\bibfnamefont {J.~G.}\ \bibnamefont {Muga}},\
  }\bibfield  {title} {\bibinfo {title} {Chapter 2 - shortcuts to
  adiabaticity},\ }in\ \href
  {https://doi.org/https://doi.org/10.1016/B978-0-12-408090-4.00002-5} {\emph
  {\bibinfo {booktitle} {Advances in Atomic, Molecular, and Optical
  Physics}}},\ \bibinfo {series} {Advances In Atomic, Molecular, and Optical
  Physics}, Vol.~\bibinfo {volume} {62},\ \bibinfo {editor} {edited by\
  \bibinfo {editor} {\bibfnamefont {E.}~\bibnamefont {Arimondo}}, \bibinfo
  {editor} {\bibfnamefont {P.~R.}\ \bibnamefont {Berman}},\ and\ \bibinfo
  {editor} {\bibfnamefont {C.~C.}\ \bibnamefont {Lin}}}\ (\bibinfo  {publisher}
  {Academic Press},\ \bibinfo {year} {2013})\ pp.\ \bibinfo {pages}
  {117--169}\BibitemShut {NoStop}%
\bibitem [{\citenamefont {Duncan}\ \emph {et~al.}(2025)\citenamefont {Duncan},
  \citenamefont {Poggi}, \citenamefont {Bukov}, \citenamefont {Zinner},\ and\
  \citenamefont {Campbell}}]{Duncan2025}%
  \BibitemOpen
  \bibfield  {author} {\bibinfo {author} {\bibfnamefont {C.~W.}\ \bibnamefont
  {Duncan}}, \bibinfo {author} {\bibfnamefont {P.~M.}\ \bibnamefont {Poggi}},
  \bibinfo {author} {\bibfnamefont {M.}~\bibnamefont {Bukov}}, \bibinfo
  {author} {\bibfnamefont {N.~T.}\ \bibnamefont {Zinner}},\ and\ \bibinfo
  {author} {\bibfnamefont {S.}~\bibnamefont {Campbell}},\ }\bibfield  {title}
  {\bibinfo {title} {{Taming quantum systems: A tutorial for using
  shortcuts-to-adiabaticity, quantum optimal control, and reinforcement
  learning}},\ }\bibfield  {journal} {\bibinfo  {journal} {arXiv}\ }\href
  {https://doi.org/10.48550/arXiv.2501.16436} {10.48550/arXiv.2501.16436}
  (\bibinfo {year} {2025}),\ \Eprint {https://arxiv.org/abs/2501.16436}
  {2501.16436} \BibitemShut {NoStop}%
\bibitem [{\citenamefont {Unanyan}\ \emph {et~al.}(1997)\citenamefont
  {Unanyan}, \citenamefont {Yatsenko}, \citenamefont {Bergmann},\ and\
  \citenamefont {Shore}}]{Unanyan1997}%
  \BibitemOpen
  \bibfield  {author} {\bibinfo {author} {\bibfnamefont {R.~G.}\ \bibnamefont
  {Unanyan}}, \bibinfo {author} {\bibfnamefont {L.~P.}\ \bibnamefont
  {Yatsenko}}, \bibinfo {author} {\bibfnamefont {K.}~\bibnamefont {Bergmann}},\
  and\ \bibinfo {author} {\bibfnamefont {B.~W.}\ \bibnamefont {Shore}},\
  }\bibfield  {title} {\bibinfo {title} {{Laser-induced adiabatic atomic
  reorientation with control of diabatic losses}},\ }\href@noop {} {\bibfield
  {journal} {\bibinfo  {journal} {Opt. Commun.}\ }\textbf {\bibinfo {volume}
  {139}},\ \bibinfo {pages} {48} (\bibinfo {year} {1997})}\BibitemShut
  {NoStop}%
\bibitem [{\citenamefont {Demirplak}\ and\ \citenamefont
  {Rice}(2003)}]{Demirplak03}%
  \BibitemOpen
  \bibfield  {author} {\bibinfo {author} {\bibfnamefont {M.}~\bibnamefont
  {Demirplak}}\ and\ \bibinfo {author} {\bibfnamefont {S.~A.}\ \bibnamefont
  {Rice}},\ }\bibfield  {title} {\bibinfo {title} {{Adiabatic Population
  Transfer with Control Fields}},\ }\href@noop {} {\bibfield  {journal}
  {\bibinfo  {journal} {J. Phys. Chem. A}\ }\textbf {\bibinfo {volume} {107}},\
  \bibinfo {pages} {9937} (\bibinfo {year} {2003})}\BibitemShut {NoStop}%
\bibitem [{\citenamefont {Demirplak}\ and\ \citenamefont
  {Rice}(2005)}]{Demirplak05}%
  \BibitemOpen
  \bibfield  {author} {\bibinfo {author} {\bibfnamefont {M.}~\bibnamefont
  {Demirplak}}\ and\ \bibinfo {author} {\bibfnamefont {S.~A.}\ \bibnamefont
  {Rice}},\ }\bibfield  {title} {\bibinfo {title} {{Assisted Adiabatic Passage
  Revisited}},\ }\href@noop {} {\bibfield  {journal} {\bibinfo  {journal} {J.
  Phys. Chem. A}\ }\textbf {\bibinfo {volume} {109}},\ \bibinfo {pages} {6838}
  (\bibinfo {year} {2005})}\BibitemShut {NoStop}%
\bibitem [{\citenamefont {Demirplak}\ and\ \citenamefont
  {Rice}(2008)}]{Demirplak08}%
  \BibitemOpen
  \bibfield  {author} {\bibinfo {author} {\bibfnamefont {M.}~\bibnamefont
  {Demirplak}}\ and\ \bibinfo {author} {\bibfnamefont {S.~A.}\ \bibnamefont
  {Rice}},\ }\bibfield  {title} {\bibinfo {title} {{On the consistency,
  extremal, and global properties of counterdiabatic fields}},\ }\href@noop {}
  {\bibfield  {journal} {\bibinfo  {journal} {J. Chem. Phys.}\ }\textbf
  {\bibinfo {volume} {129}},\ \bibinfo {pages} {154111} (\bibinfo {year}
  {2008})}\BibitemShut {NoStop}%
\bibitem [{\citenamefont {Berry}(2009)}]{Berry09}%
  \BibitemOpen
  \bibfield  {author} {\bibinfo {author} {\bibfnamefont {M.~V.}\ \bibnamefont
  {Berry}},\ }\bibfield  {title} {\bibinfo {title} {{Transitionless quantum
  driving}},\ }\href@noop {} {\bibfield  {journal} {\bibinfo  {journal} {J.
  Phys. A: Math. Theor.}\ }\textbf {\bibinfo {volume} {42}},\ \bibinfo {pages}
  {365303} (\bibinfo {year} {2009})}\BibitemShut {NoStop}%
\bibitem [{\citenamefont {Del~Campo}(2013)}]{prl2013_adolfo}%
  \BibitemOpen
  \bibfield  {author} {\bibinfo {author} {\bibfnamefont {A.}~\bibnamefont
  {Del~Campo}},\ }\bibfield  {title} {\bibinfo {title} {Shortcuts to
  adiabaticity by counterdiabatic driving},\ }\href@noop {} {\bibfield
  {journal} {\bibinfo  {journal} {Physical Review Letters}\ }\textbf {\bibinfo
  {volume} {111}},\ \bibinfo {pages} {100502} (\bibinfo {year}
  {2013})}\BibitemShut {NoStop}%
\bibitem [{\citenamefont {Del~Campo}\ \emph {et~al.}(2014)\citenamefont
  {Del~Campo}, \citenamefont {Goold},\ and\ \citenamefont
  {Paternostro}}]{Campo14}%
  \BibitemOpen
  \bibfield  {author} {\bibinfo {author} {\bibfnamefont {A.}~\bibnamefont
  {Del~Campo}}, \bibinfo {author} {\bibfnamefont {J.}~\bibnamefont {Goold}},\
  and\ \bibinfo {author} {\bibfnamefont {M.}~\bibnamefont {Paternostro}},\
  }\bibfield  {title} {\bibinfo {title} {{More bang for your buck:
  Super-adiabatic quantum engines}},\ }\href
  {https://doi.org/10.1038/srep06208} {\bibfield  {journal} {\bibinfo
  {journal} {Sci. Rep.}\ }\textbf {\bibinfo {volume} {4}},\ \bibinfo {pages}
  {1} (\bibinfo {year} {2014})}\BibitemShut {NoStop}%
\bibitem [{\citenamefont {Deng}\ \emph {et~al.}(2013)\citenamefont {Deng},
  \citenamefont {Wang}, \citenamefont {Liu}, \citenamefont
  {H{\ifmmode\ddot{a}\else\"{a}\fi}nggi},\ and\ \citenamefont {Gong}}]{Deng13}%
  \BibitemOpen
  \bibfield  {author} {\bibinfo {author} {\bibfnamefont {J.}~\bibnamefont
  {Deng}}, \bibinfo {author} {\bibfnamefont {Q.-h.}\ \bibnamefont {Wang}},
  \bibinfo {author} {\bibfnamefont {Z.}~\bibnamefont {Liu}}, \bibinfo {author}
  {\bibfnamefont {P.}~\bibnamefont {H{\ifmmode\ddot{a}\else\"{a}\fi}nggi}},\
  and\ \bibinfo {author} {\bibfnamefont {J.}~\bibnamefont {Gong}},\ }\bibfield
  {title} {\bibinfo {title} {{Boosting work characteristics and overall
  heat-engine performance via shortcuts to adiabaticity: Quantum and classical
  systems}},\ }\href {https://doi.org/10.1103/PhysRevE.88.062122} {\bibfield
  {journal} {\bibinfo  {journal} {Phys. Rev. E}\ }\textbf {\bibinfo {volume}
  {88}},\ \bibinfo {pages} {062122} (\bibinfo {year} {2013})}\BibitemShut
  {NoStop}%
\bibitem [{\citenamefont {Beau}\ \emph {et~al.}(2016)\citenamefont {Beau},
  \citenamefont {Jaramillo},\ and\ \citenamefont {Del~Campo}}]{Beau16}%
  \BibitemOpen
  \bibfield  {author} {\bibinfo {author} {\bibfnamefont {M.}~\bibnamefont
  {Beau}}, \bibinfo {author} {\bibfnamefont {J.}~\bibnamefont {Jaramillo}},\
  and\ \bibinfo {author} {\bibfnamefont {A.}~\bibnamefont {Del~Campo}},\
  }\bibfield  {title} {\bibinfo {title} {{Scaling-Up Quantum Heat Engines
  Efficiently via Shortcuts to Adiabaticity}},\ }\href
  {https://doi.org/10.3390/e18050168} {\bibfield  {journal} {\bibinfo
  {journal} {Entropy}\ }\textbf {\bibinfo {volume} {18}},\ \bibinfo {pages}
  {168} (\bibinfo {year} {2016})}\BibitemShut {NoStop}%
\bibitem [{\citenamefont {Abah}\ and\ \citenamefont
  {Paternostro}(2019{\natexlab{a}})}]{Abah19}%
  \BibitemOpen
  \bibfield  {author} {\bibinfo {author} {\bibfnamefont {O.}~\bibnamefont
  {Abah}}\ and\ \bibinfo {author} {\bibfnamefont {M.}~\bibnamefont
  {Paternostro}},\ }\bibfield  {title} {\bibinfo {title}
  {{Shortcut-to-adiabaticity Otto engine: A twist to finite-time
  thermodynamics}},\ }\href {https://doi.org/10.1103/PhysRevE.99.022110}
  {\bibfield  {journal} {\bibinfo  {journal} {Phys. Rev. E}\ }\textbf {\bibinfo
  {volume} {99}},\ \bibinfo {pages} {022110} (\bibinfo {year}
  {2019}{\natexlab{a}})}\BibitemShut {NoStop}%
\bibitem [{\citenamefont {Hartmann}\ \emph
  {et~al.}(2020{\natexlab{a}})\citenamefont {Hartmann}, \citenamefont
  {Mukherjee}, \citenamefont {Niedenzu},\ and\ \citenamefont
  {Lechner}}]{Hartmann20}%
  \BibitemOpen
  \bibfield  {author} {\bibinfo {author} {\bibfnamefont {A.}~\bibnamefont
  {Hartmann}}, \bibinfo {author} {\bibfnamefont {V.}~\bibnamefont {Mukherjee}},
  \bibinfo {author} {\bibfnamefont {W.}~\bibnamefont {Niedenzu}},\ and\
  \bibinfo {author} {\bibfnamefont {W.}~\bibnamefont {Lechner}},\ }\bibfield
  {title} {\bibinfo {title} {{Many-body quantum heat engines with shortcuts to
  adiabaticity}},\ }\href {https://doi.org/10.1103/PhysRevResearch.2.023145}
  {\bibfield  {journal} {\bibinfo  {journal} {Phys. Rev. Res.}\ }\textbf
  {\bibinfo {volume} {2}},\ \bibinfo {pages} {023145} (\bibinfo {year}
  {2020}{\natexlab{a}})}\BibitemShut {NoStop}%
\bibitem [{\citenamefont {Dann}\ \emph {et~al.}(2020)\citenamefont {Dann},
  \citenamefont {Kosloff},\ and\ \citenamefont {Salamon}}]{Dann20}%
  \BibitemOpen
  \bibfield  {author} {\bibinfo {author} {\bibfnamefont {R.}~\bibnamefont
  {Dann}}, \bibinfo {author} {\bibfnamefont {R.}~\bibnamefont {Kosloff}},\ and\
  \bibinfo {author} {\bibfnamefont {P.}~\bibnamefont {Salamon}},\ }\bibfield
  {title} {\bibinfo {title} {{Quantum Finite-Time Thermodynamics: Insight from
  a Single Qubit Engine}},\ }\href {https://doi.org/10.3390/e22111255}
  {\bibfield  {journal} {\bibinfo  {journal} {Entropy}\ }\textbf {\bibinfo
  {volume} {22}},\ \bibinfo {pages} {1255} (\bibinfo {year}
  {2020})}\BibitemShut {NoStop}%
\bibitem [{\citenamefont {Dann}\ and\ \citenamefont
  {Kosloff}(2020)}]{Dann2020Jan}%
  \BibitemOpen
  \bibfield  {author} {\bibinfo {author} {\bibfnamefont {R.}~\bibnamefont
  {Dann}}\ and\ \bibinfo {author} {\bibfnamefont {R.}~\bibnamefont {Kosloff}},\
  }\bibfield  {title} {\bibinfo {title} {{Quantum signatures in the quantum
  Carnot cycle}},\ }\href {https://doi.org/10.1088/1367-2630/ab6876} {\bibfield
   {journal} {\bibinfo  {journal} {New J. Phys.}\ }\textbf {\bibinfo {volume}
  {22}},\ \bibinfo {pages} {013055} (\bibinfo {year} {2020})}\BibitemShut
  {NoStop}%
\bibitem [{\citenamefont {Deng}\ \emph {et~al.}(2018)\citenamefont {Deng},
  \citenamefont {Chenu}, \citenamefont {Diao}, \citenamefont {Li},
  \citenamefont {Yu}, \citenamefont {Coulamy}, \citenamefont {del Campo},\ and\
  \citenamefont {Wu}}]{Deng18}%
  \BibitemOpen
  \bibfield  {author} {\bibinfo {author} {\bibfnamefont {S.}~\bibnamefont
  {Deng}}, \bibinfo {author} {\bibfnamefont {A.}~\bibnamefont {Chenu}},
  \bibinfo {author} {\bibfnamefont {P.}~\bibnamefont {Diao}}, \bibinfo {author}
  {\bibfnamefont {F.}~\bibnamefont {Li}}, \bibinfo {author} {\bibfnamefont
  {S.}~\bibnamefont {Yu}}, \bibinfo {author} {\bibfnamefont {I.}~\bibnamefont
  {Coulamy}}, \bibinfo {author} {\bibfnamefont {A.}~\bibnamefont {del Campo}},\
  and\ \bibinfo {author} {\bibfnamefont {H.}~\bibnamefont {Wu}},\ }\bibfield
  {title} {\bibinfo {title} {{Superadiabatic quantum friction suppression in
  finite-time thermodynamics}},\ }\bibfield  {journal} {\bibinfo  {journal}
  {Sci. Adv.}\ }\textbf {\bibinfo {volume} {4}},\ \href
  {https://doi.org/10.1126/sciadv.aar5909} {10.1126/sciadv.aar5909} (\bibinfo
  {year} {2018})\BibitemShut {NoStop}%
\bibitem [{\citenamefont {Kosloff}\ and\ \citenamefont
  {Feldmann}(2002)}]{Kosloff02}%
  \BibitemOpen
  \bibfield  {author} {\bibinfo {author} {\bibfnamefont {R.}~\bibnamefont
  {Kosloff}}\ and\ \bibinfo {author} {\bibfnamefont {T.}~\bibnamefont
  {Feldmann}},\ }\bibfield  {title} {\bibinfo {title} {{Discrete four-stroke
  quantum heat engine exploring the origin of friction}},\ }\href@noop {}
  {\bibfield  {journal} {\bibinfo  {journal} {Phys. Rev. E}\ }\textbf {\bibinfo
  {volume} {65}},\ \bibinfo {pages} {055102} (\bibinfo {year}
  {2002})}\BibitemShut {NoStop}%
\bibitem [{\citenamefont {Feldmann}\ and\ \citenamefont
  {Kosloff}(2003)}]{Feldmann03}%
  \BibitemOpen
  \bibfield  {author} {\bibinfo {author} {\bibfnamefont {T.}~\bibnamefont
  {Feldmann}}\ and\ \bibinfo {author} {\bibfnamefont {R.}~\bibnamefont
  {Kosloff}},\ }\bibfield  {title} {\bibinfo {title} {{Quantum four-stroke heat
  engine: Thermodynamic observables in a model with intrinsic friction}},\
  }\href@noop {} {\bibfield  {journal} {\bibinfo  {journal} {Phys. Rev. E}\
  }\textbf {\bibinfo {volume} {68}},\ \bibinfo {pages} {016101} (\bibinfo
  {year} {2003})}\BibitemShut {NoStop}%
\bibitem [{\citenamefont {Feldmann}\ and\ \citenamefont
  {Kosloff}(2004)}]{Feldmann04}%
  \BibitemOpen
  \bibfield  {author} {\bibinfo {author} {\bibfnamefont {T.}~\bibnamefont
  {Feldmann}}\ and\ \bibinfo {author} {\bibfnamefont {R.}~\bibnamefont
  {Kosloff}},\ }\bibfield  {title} {\bibinfo {title} {{Characteristics of the
  limit cycle of a reciprocating quantum heat engine}},\ }\href@noop {}
  {\bibfield  {journal} {\bibinfo  {journal} {Phys. Rev. E}\ }\textbf {\bibinfo
  {volume} {70}},\ \bibinfo {pages} {046110} (\bibinfo {year}
  {2004})}\BibitemShut {NoStop}%
\bibitem [{\citenamefont {Hegade}\ \emph {et~al.}(2021)\citenamefont {Hegade},
  \citenamefont {Paul}, \citenamefont {Ding}, \citenamefont {Sanz},
  \citenamefont {Albarr{\ifmmode\acute{a}\else\'{a}\fi}n-Arriagada},
  \citenamefont {Solano},\ and\ \citenamefont {Chen}}]{Hegade21}%
  \BibitemOpen
  \bibfield  {author} {\bibinfo {author} {\bibfnamefont {N.~N.}\ \bibnamefont
  {Hegade}}, \bibinfo {author} {\bibfnamefont {K.}~\bibnamefont {Paul}},
  \bibinfo {author} {\bibfnamefont {Y.}~\bibnamefont {Ding}}, \bibinfo {author}
  {\bibfnamefont {M.}~\bibnamefont {Sanz}}, \bibinfo {author} {\bibfnamefont
  {F.}~\bibnamefont {Albarr{\ifmmode\acute{a}\else\'{a}\fi}n-Arriagada}},
  \bibinfo {author} {\bibfnamefont {E.}~\bibnamefont {Solano}},\ and\ \bibinfo
  {author} {\bibfnamefont {X.}~\bibnamefont {Chen}},\ }\bibfield  {title}
  {\bibinfo {title} {{Shortcuts to Adiabaticity in Digitized Adiabatic Quantum
  Computing}},\ }\href {https://doi.org/10.1103/PhysRevApplied.15.024038}
  {\bibfield  {journal} {\bibinfo  {journal} {Phys. Rev. Appl.}\ }\textbf
  {\bibinfo {volume} {15}},\ \bibinfo {pages} {024038} (\bibinfo {year}
  {2021})}\BibitemShut {NoStop}%
\bibitem [{\citenamefont {Chen}\ \emph {et~al.}(2021)\citenamefont {Chen},
  \citenamefont {Qin}, \citenamefont {Wang}, \citenamefont {Miranowicz},\ and\
  \citenamefont {Nori}}]{Chen21}%
  \BibitemOpen
  \bibfield  {author} {\bibinfo {author} {\bibfnamefont {Y.-H.}\ \bibnamefont
  {Chen}}, \bibinfo {author} {\bibfnamefont {W.}~\bibnamefont {Qin}}, \bibinfo
  {author} {\bibfnamefont {X.}~\bibnamefont {Wang}}, \bibinfo {author}
  {\bibfnamefont {A.}~\bibnamefont {Miranowicz}},\ and\ \bibinfo {author}
  {\bibfnamefont {F.}~\bibnamefont {Nori}},\ }\bibfield  {title} {\bibinfo
  {title} {{Shortcuts to Adiabaticity for the Quantum Rabi Model: Efficient
  Generation of Giant Entangled Cat States via Parametric Amplification}},\
  }\href {https://doi.org/10.1103/PhysRevLett.126.023602} {\bibfield  {journal}
  {\bibinfo  {journal} {Phys. Rev. Lett.}\ }\textbf {\bibinfo {volume} {126}},\
  \bibinfo {pages} {023602} (\bibinfo {year} {2021})}\BibitemShut {NoStop}%
\bibitem [{\citenamefont {Santos}\ \emph {et~al.}(2020)\citenamefont {Santos},
  \citenamefont {Nicotina}, \citenamefont {Souza}, \citenamefont {Sarthour},
  \citenamefont {Oliveira},\ and\ \citenamefont {Sarandy}}]{Santos20}%
  \BibitemOpen
  \bibfield  {author} {\bibinfo {author} {\bibfnamefont {A.~C.}\ \bibnamefont
  {Santos}}, \bibinfo {author} {\bibfnamefont {A.}~\bibnamefont {Nicotina}},
  \bibinfo {author} {\bibfnamefont {A.~M.}\ \bibnamefont {Souza}}, \bibinfo
  {author} {\bibfnamefont {R.~S.}\ \bibnamefont {Sarthour}}, \bibinfo {author}
  {\bibfnamefont {I.~S.}\ \bibnamefont {Oliveira}},\ and\ \bibinfo {author}
  {\bibfnamefont {M.~S.}\ \bibnamefont {Sarandy}},\ }\bibfield  {title}
  {\bibinfo {title} {{Optimizing NMR quantum information processing via
  generalized transitionless quantum driving}},\ }\href
  {https://doi.org/10.1209/0295-5075/129/30008} {\bibfield  {journal} {\bibinfo
   {journal} {Europhys. Lett.}\ }\textbf {\bibinfo {volume} {129}},\ \bibinfo
  {pages} {30008} (\bibinfo {year} {2020})}\BibitemShut {NoStop}%
\bibitem [{\citenamefont {Wang}\ \emph {et~al.}(2015)\citenamefont {Wang},
  \citenamefont {Allegra}, \citenamefont {Jacobs}, \citenamefont {Lloyd},
  \citenamefont {Lupo},\ and\ \citenamefont {Mohseni}}]{Wang15}%
  \BibitemOpen
  \bibfield  {author} {\bibinfo {author} {\bibfnamefont {X.}~\bibnamefont
  {Wang}}, \bibinfo {author} {\bibfnamefont {M.}~\bibnamefont {Allegra}},
  \bibinfo {author} {\bibfnamefont {K.}~\bibnamefont {Jacobs}}, \bibinfo
  {author} {\bibfnamefont {S.}~\bibnamefont {Lloyd}}, \bibinfo {author}
  {\bibfnamefont {C.}~\bibnamefont {Lupo}},\ and\ \bibinfo {author}
  {\bibfnamefont {M.}~\bibnamefont {Mohseni}},\ }\bibfield  {title} {\bibinfo
  {title} {{Quantum Brachistochrone Curves as Geodesics: Obtaining Accurate
  Minimum-Time Protocols for the Control of Quantum Systems}},\ }\href
  {https://doi.org/10.1103/PhysRevLett.114.170501} {\bibfield  {journal}
  {\bibinfo  {journal} {Phys. Rev. Lett.}\ }\textbf {\bibinfo {volume} {114}},\
  \bibinfo {pages} {170501} (\bibinfo {year} {2015})}\BibitemShut {NoStop}%
\bibitem [{\citenamefont {Koike}(2022)}]{Koike2022}%
  \BibitemOpen
  \bibfield  {author} {\bibinfo {author} {\bibfnamefont {T.}~\bibnamefont
  {Koike}},\ }\bibfield  {title} {\bibinfo {title} {{Quantum
  brachistochrone}},\ }\bibfield  {journal} {\bibinfo  {journal} {Philos.
  Trans. Royal Soc. A}\ }\textbf {\bibinfo {volume} {380}},\ \href
  {https://doi.org/10.1098/rsta.2021.0273} {10.1098/rsta.2021.0273} (\bibinfo
  {year} {2022})\BibitemShut {NoStop}%
\bibitem [{\citenamefont {Liberzon}(2012)}]{liberzon-book}%
  \BibitemOpen
  \bibfield  {author} {\bibinfo {author} {\bibfnamefont {D.}~\bibnamefont
  {Liberzon}},\ }\href@noop {} {\emph {\bibinfo {title} {Calculus of variations
  and optimal control theory}}}\ (\bibinfo  {publisher} {Princeton University
  Press, Princeton, NJ},\ \bibinfo {year} {2012})\ pp.\ \bibinfo {pages}
  {xviii+235}\BibitemShut {NoStop}%
\bibitem [{\citenamefont {D'Alessandro}(2008)}]{dalessandro-book}%
  \BibitemOpen
  \bibfield  {author} {\bibinfo {author} {\bibfnamefont {D.}~\bibnamefont
  {D'Alessandro}},\ }\href@noop {} {\emph {\bibinfo {title} {{Introduction to
  quantum control and dynamics.}}}}\ (\bibinfo  {publisher} {{Applied
  Mathematics and Nonlinear Science Series. Boca Raton, FL: Chapman,
  Hall/CRC.}},\ \bibinfo {year} {2008})\BibitemShut {NoStop}%
\bibitem [{\citenamefont {Kirk}(2004)}]{kirk2004optimal}%
  \BibitemOpen
  \bibfield  {author} {\bibinfo {author} {\bibfnamefont {D.~E.}\ \bibnamefont
  {Kirk}},\ }\href@noop {} {\emph {\bibinfo {title} {Optimal control theory: an
  introduction}}}\ (\bibinfo  {publisher} {Courier Corporation, New York},\
  \bibinfo {year} {2004})\BibitemShut {NoStop}%
\bibitem [{\citenamefont {Liu}\ \emph {et~al.}(2023)\citenamefont {Liu},
  \citenamefont {Sugny}, \citenamefont {Chen},\ and\ \citenamefont
  {Gu{\ifmmode\acute{e}\else\'{e}\fi}rin}}]{Liu2023}%
  \BibitemOpen
  \bibfield  {author} {\bibinfo {author} {\bibfnamefont {K.}~\bibnamefont
  {Liu}}, \bibinfo {author} {\bibfnamefont {D.}~\bibnamefont {Sugny}}, \bibinfo
  {author} {\bibfnamefont {X.}~\bibnamefont {Chen}},\ and\ \bibinfo {author}
  {\bibfnamefont {S.}~\bibnamefont {Gu{\ifmmode\acute{e}\else\'{e}\fi}rin}},\
  }\bibfield  {title} {\bibinfo {title} {{Optimal Pulse Design for
  Dissipative-Stimulated Raman Exact Passage}},\ }\href
  {https://doi.org/10.3390/e25050790} {\bibfield  {journal} {\bibinfo
  {journal} {Entropy}\ }\textbf {\bibinfo {volume} {25}},\ \bibinfo {pages}
  {790} (\bibinfo {year} {2023})}\BibitemShut {NoStop}%
\bibitem [{\citenamefont {Dridi}\ \emph {et~al.}(2020)\citenamefont {Dridi},
  \citenamefont {Liu},\ and\ \citenamefont
  {Gu{\ifmmode\acute{e}\else\'{e}\fi}rin}}]{Dridi2020}%
  \BibitemOpen
  \bibfield  {author} {\bibinfo {author} {\bibfnamefont {G.}~\bibnamefont
  {Dridi}}, \bibinfo {author} {\bibfnamefont {K.}~\bibnamefont {Liu}},\ and\
  \bibinfo {author} {\bibfnamefont {S.}~\bibnamefont
  {Gu{\ifmmode\acute{e}\else\'{e}\fi}rin}},\ }\bibfield  {title} {\bibinfo
  {title} {{Optimal Robust Quantum Control by Inverse Geometric
  Optimization}},\ }\href {https://doi.org/10.1103/PhysRevLett.125.250403}
  {\bibfield  {journal} {\bibinfo  {journal} {Phys. Rev. Lett.}\ }\textbf
  {\bibinfo {volume} {125}},\ \bibinfo {pages} {250403} (\bibinfo {year}
  {2020})}\BibitemShut {NoStop}%
\bibitem [{\citenamefont {Harutyunyan}\ \emph {et~al.}(2023)\citenamefont
  {Harutyunyan}, \citenamefont {Holweck}, \citenamefont {Sugny},\ and\
  \citenamefont {Gu\'erin}}]{meri2023}%
  \BibitemOpen
  \bibfield  {author} {\bibinfo {author} {\bibfnamefont {M.}~\bibnamefont
  {Harutyunyan}}, \bibinfo {author} {\bibfnamefont {F.}~\bibnamefont
  {Holweck}}, \bibinfo {author} {\bibfnamefont {D.}~\bibnamefont {Sugny}},\
  and\ \bibinfo {author} {\bibfnamefont {S.}~\bibnamefont {Gu\'erin}},\
  }\bibfield  {title} {\bibinfo {title} {Digital optimal robust control},\
  }\href {https://doi.org/10.1103/PhysRevLett.131.200801} {\bibfield  {journal}
  {\bibinfo  {journal} {Phys. Rev. Lett.}\ }\textbf {\bibinfo {volume} {131}},\
  \bibinfo {pages} {200801} (\bibinfo {year} {2023})}\BibitemShut {NoStop}%
\bibitem [{\citenamefont {Van~Damme}\ \emph {et~al.}(2017)\citenamefont
  {Van~Damme}, \citenamefont {Ansel}, \citenamefont {Glaser},\ and\
  \citenamefont {Sugny}}]{vandamme2017}%
  \BibitemOpen
  \bibfield  {author} {\bibinfo {author} {\bibfnamefont {L.}~\bibnamefont
  {Van~Damme}}, \bibinfo {author} {\bibfnamefont {Q.}~\bibnamefont {Ansel}},
  \bibinfo {author} {\bibfnamefont {S.~J.}\ \bibnamefont {Glaser}},\ and\
  \bibinfo {author} {\bibfnamefont {D.}~\bibnamefont {Sugny}},\ }\bibfield
  {title} {\bibinfo {title} {Robust optimal control of two-level quantum
  systems},\ }\href {https://doi.org/10.1103/PhysRevA.95.063403} {\bibfield
  {journal} {\bibinfo  {journal} {Phys. Rev. A}\ }\textbf {\bibinfo {volume}
  {95}},\ \bibinfo {pages} {063403} (\bibinfo {year} {2017})}\BibitemShut
  {NoStop}%
\bibitem [{\citenamefont {Carolan}\ \emph {et~al.}(2023)\citenamefont
  {Carolan}, \citenamefont {{\ifmmode\mbox{\c{C}}\else\c{C}\fi}akmak},\ and\
  \citenamefont {Campbell}}]{Carolan23}%
  \BibitemOpen
  \bibfield  {author} {\bibinfo {author} {\bibfnamefont {E.}~\bibnamefont
  {Carolan}}, \bibinfo {author} {\bibfnamefont {B.}~\bibnamefont
  {{\ifmmode\mbox{\c{C}}\else\c{C}\fi}akmak}},\ and\ \bibinfo {author}
  {\bibfnamefont {S.}~\bibnamefont {Campbell}},\ }\bibfield  {title} {\bibinfo
  {title} {{Robustness of controlled Hamiltonian approaches to unitary quantum
  gates}},\ }\href {https://doi.org/10.1103/PhysRevA.108.022423} {\bibfield
  {journal} {\bibinfo  {journal} {Phys. Rev. A}\ }\textbf {\bibinfo {volume}
  {108}},\ \bibinfo {pages} {022423} (\bibinfo {year} {2023})}\BibitemShut
  {NoStop}%
\bibitem [{\citenamefont {Vitanov}\ \emph {et~al.}(2017)\citenamefont
  {Vitanov}, \citenamefont {Rangelov}, \citenamefont {Shore},\ and\
  \citenamefont {Bergmann}}]{Vitanov}%
  \BibitemOpen
  \bibfield  {author} {\bibinfo {author} {\bibfnamefont {N.~V.}\ \bibnamefont
  {Vitanov}}, \bibinfo {author} {\bibfnamefont {A.~A.}\ \bibnamefont
  {Rangelov}}, \bibinfo {author} {\bibfnamefont {B.~W.}\ \bibnamefont
  {Shore}},\ and\ \bibinfo {author} {\bibfnamefont {K.}~\bibnamefont
  {Bergmann}},\ }\bibfield  {title} {\bibinfo {title} {Stimulated raman
  adiabatic passage in physics, chemistry, and beyond},\ }\href
  {https://doi.org/10.1103/RevModPhys.89.015006} {\bibfield  {journal}
  {\bibinfo  {journal} {Rev. Mod. Phys.}\ }\textbf {\bibinfo {volume} {89}},\
  \bibinfo {pages} {015006} (\bibinfo {year} {2017})}\BibitemShut {NoStop}%
\bibitem [{\citenamefont {Born}\ and\ \citenamefont {Fock}(1928)}]{Born1928}%
  \BibitemOpen
  \bibfield  {author} {\bibinfo {author} {\bibfnamefont {M.}~\bibnamefont
  {Born}}\ and\ \bibinfo {author} {\bibfnamefont {V.}~\bibnamefont {Fock}},\
  }\bibfield  {title} {\bibinfo {title} {{Beweis des Adiabatensatzes}},\ }\href
  {https://doi.org/10.1007/BF01343193} {\bibfield  {journal} {\bibinfo
  {journal} {Z. Phys.}\ }\textbf {\bibinfo {volume} {51}},\ \bibinfo {pages}
  {165} (\bibinfo {year} {1928})}\BibitemShut {NoStop}%
\bibitem [{\citenamefont {Abah}\ and\ \citenamefont
  {Paternostro}(2019{\natexlab{b}})}]{Abah2019}%
  \BibitemOpen
  \bibfield  {author} {\bibinfo {author} {\bibfnamefont {O.}~\bibnamefont
  {Abah}}\ and\ \bibinfo {author} {\bibfnamefont {M.}~\bibnamefont
  {Paternostro}},\ }\bibfield  {title} {\bibinfo {title}
  {{Shortcut-to-adiabaticity Otto engine: A twist to finite-time
  thermodynamics}},\ }\href {https://doi.org/10.1103/PhysRevE.99.022110}
  {\bibfield  {journal} {\bibinfo  {journal} {Phys. Rev. E}\ }\textbf {\bibinfo
  {volume} {99}},\ \bibinfo {pages} {022110} (\bibinfo {year}
  {2019}{\natexlab{b}})}\BibitemShut {NoStop}%
\bibitem [{\citenamefont {Hartmann}\ \emph
  {et~al.}(2020{\natexlab{b}})\citenamefont {Hartmann}, \citenamefont
  {Mukherjee}, \citenamefont {Niedenzu},\ and\ \citenamefont
  {Lechner}}]{Hartmann2020}%
  \BibitemOpen
  \bibfield  {author} {\bibinfo {author} {\bibfnamefont {A.}~\bibnamefont
  {Hartmann}}, \bibinfo {author} {\bibfnamefont {V.}~\bibnamefont {Mukherjee}},
  \bibinfo {author} {\bibfnamefont {W.}~\bibnamefont {Niedenzu}},\ and\
  \bibinfo {author} {\bibfnamefont {W.}~\bibnamefont {Lechner}},\ }\bibfield
  {title} {\bibinfo {title} {{Many-body quantum heat engines with shortcuts to
  adiabaticity}},\ }\href {https://doi.org/10.1103/PhysRevResearch.2.023145}
  {\bibfield  {journal} {\bibinfo  {journal} {Phys. Rev. Res.}\ }\textbf
  {\bibinfo {volume} {2}},\ \bibinfo {pages} {023145} (\bibinfo {year}
  {2020}{\natexlab{b}})}\BibitemShut {NoStop}%
\bibitem [{\citenamefont {Daems}\ \emph {et~al.}(2008)\citenamefont {Daems},
  \citenamefont {Gu{\ifmmode\acute{e}\else\'{e}\fi}rin},\ and\ \citenamefont
  {Cerf}}]{Daems2008}%
  \BibitemOpen
  \bibfield  {author} {\bibinfo {author} {\bibfnamefont {D.}~\bibnamefont
  {Daems}}, \bibinfo {author} {\bibfnamefont {S.}~\bibnamefont
  {Gu{\ifmmode\acute{e}\else\'{e}\fi}rin}},\ and\ \bibinfo {author}
  {\bibfnamefont {N.~J.}\ \bibnamefont {Cerf}},\ }\bibfield  {title} {\bibinfo
  {title} {{Quantum search by parallel eigenvalue adiabatic passage}},\ }\href
  {https://doi.org/10.1103/PhysRevA.78.042322} {\bibfield  {journal} {\bibinfo
  {journal} {Phys. Rev. A}\ }\textbf {\bibinfo {volume} {78}},\ \bibinfo
  {pages} {042322} (\bibinfo {year} {2008})}\BibitemShut {NoStop}%
\bibitem [{\citenamefont {Allahverdyan}\ and\ \citenamefont
  {Nieuwenhuizen}(2005)}]{Allahverdyan05}%
  \BibitemOpen
  \bibfield  {author} {\bibinfo {author} {\bibfnamefont {A.~E.}\ \bibnamefont
  {Allahverdyan}}\ and\ \bibinfo {author} {\bibfnamefont {{\relax Th}.~M.}\
  \bibnamefont {Nieuwenhuizen}},\ }\bibfield  {title} {\bibinfo {title}
  {{Minimal work principle: Proof and counterexamples}},\ }\href
  {https://doi.org/10.1103/PhysRevE.71.046107} {\bibfield  {journal} {\bibinfo
  {journal} {Phys. Rev. E}\ }\textbf {\bibinfo {volume} {71}},\ \bibinfo
  {pages} {046107} (\bibinfo {year} {2005})}\BibitemShut {NoStop}%
\bibitem [{\citenamefont {Albash}\ \emph {et~al.}(2012)\citenamefont {Albash},
  \citenamefont {Boixo}, \citenamefont {Lidar},\ and\ \citenamefont
  {Zanardi}}]{Albash12}%
  \BibitemOpen
  \bibfield  {author} {\bibinfo {author} {\bibfnamefont {T.}~\bibnamefont
  {Albash}}, \bibinfo {author} {\bibfnamefont {S.}~\bibnamefont {Boixo}},
  \bibinfo {author} {\bibfnamefont {D.~A.}\ \bibnamefont {Lidar}},\ and\
  \bibinfo {author} {\bibfnamefont {P.}~\bibnamefont {Zanardi}},\ }\bibfield
  {title} {\bibinfo {title} {{Quantum adiabatic Markovian master equations}},\
  }\href {https://doi.org/10.1088/1367-2630/14/12/123016} {\bibfield  {journal}
  {\bibinfo  {journal} {New J. Phys.}\ }\textbf {\bibinfo {volume} {14}},\
  \bibinfo {pages} {123016} (\bibinfo {year} {2012})}\BibitemShut {NoStop}%
\bibitem [{\citenamefont {Moutinho}\ \emph {et~al.}(2023)\citenamefont
  {Moutinho}, \citenamefont {Pezzutto}, \citenamefont {Pratapsi}, \citenamefont
  {da~Silva}, \citenamefont {De~Franceschi}, \citenamefont {Bose},
  \citenamefont {Costa},\ and\ \citenamefont {Omar}}]{Moutinho2023}%
  \BibitemOpen
  \bibfield  {author} {\bibinfo {author} {\bibfnamefont {J.~P.}\ \bibnamefont
  {Moutinho}}, \bibinfo {author} {\bibfnamefont {M.}~\bibnamefont {Pezzutto}},
  \bibinfo {author} {\bibfnamefont {S.~S.}\ \bibnamefont {Pratapsi}}, \bibinfo
  {author} {\bibfnamefont {F.~F.}\ \bibnamefont {da~Silva}}, \bibinfo {author}
  {\bibfnamefont {S.}~\bibnamefont {De~Franceschi}}, \bibinfo {author}
  {\bibfnamefont {S.}~\bibnamefont {Bose}}, \bibinfo {author} {\bibfnamefont
  {A.~T.}\ \bibnamefont {Costa}},\ and\ \bibinfo {author} {\bibfnamefont
  {Y.}~\bibnamefont {Omar}},\ }\bibfield  {title} {\bibinfo {title} {{Quantum
  Dynamics for Energetic Advantage in a Charge-Based Classical Full Adder}},\
  }\href {https://doi.org/10.1103/PRXEnergy.2.033002} {\bibfield  {journal}
  {\bibinfo  {journal} {PRX Energy}\ }\textbf {\bibinfo {volume} {2}},\
  \bibinfo {pages} {033002} (\bibinfo {year} {2023})}\BibitemShut {NoStop}%
\bibitem [{\citenamefont {G{\ifmmode\acute{o}\else\'{o}\fi}is}\ \emph
  {et~al.}(2024)\citenamefont {G{\ifmmode\acute{o}\else\'{o}\fi}is},
  \citenamefont {Pezzutto},\ and\ \citenamefont {Omar}}]{Gois2024}%
  \BibitemOpen
  \bibfield  {author} {\bibinfo {author} {\bibfnamefont {F.}~\bibnamefont
  {G{\ifmmode\acute{o}\else\'{o}\fi}is}}, \bibinfo {author} {\bibfnamefont
  {M.}~\bibnamefont {Pezzutto}},\ and\ \bibinfo {author} {\bibfnamefont
  {Y.}~\bibnamefont {Omar}},\ }\bibfield  {title} {\bibinfo {title} {{Towards
  Energetic Quantum Advantage in Trapped-Ion Quantum Computation}},\ }\bibfield
   {journal} {\bibinfo  {journal} {arXiv}\ }\href
  {https://doi.org/10.48550/arXiv.2404.11572} {10.48550/arXiv.2404.11572}
  (\bibinfo {year} {2024}),\ \Eprint {https://arxiv.org/abs/2404.11572}
  {2404.11572} \BibitemShut {NoStop}%
\bibitem [{\citenamefont {Campbell}\ and\ \citenamefont
  {Deffner}(2017)}]{Campbell2017}%
  \BibitemOpen
  \bibfield  {author} {\bibinfo {author} {\bibfnamefont {S.}~\bibnamefont
  {Campbell}}\ and\ \bibinfo {author} {\bibfnamefont {S.}~\bibnamefont
  {Deffner}},\ }\bibfield  {title} {\bibinfo {title} {{Trade-Off Between Speed
  and Cost in Shortcuts to Adiabaticity}},\ }\href
  {https://doi.org/10.1103/PhysRevLett.118.100601} {\bibfield  {journal}
  {\bibinfo  {journal} {Phys. Rev. Lett.}\ }\textbf {\bibinfo {volume} {118}},\
  \bibinfo {pages} {100601} (\bibinfo {year} {2017})}\BibitemShut {NoStop}%
\bibitem [{\citenamefont {Zheng}\ \emph {et~al.}(2016)\citenamefont {Zheng},
  \citenamefont {Campbell}, \citenamefont {De~Chiara},\ and\ \citenamefont
  {Poletti}}]{Zheng2016}%
  \BibitemOpen
  \bibfield  {author} {\bibinfo {author} {\bibfnamefont {Y.}~\bibnamefont
  {Zheng}}, \bibinfo {author} {\bibfnamefont {S.}~\bibnamefont {Campbell}},
  \bibinfo {author} {\bibfnamefont {G.}~\bibnamefont {De~Chiara}},\ and\
  \bibinfo {author} {\bibfnamefont {D.}~\bibnamefont {Poletti}},\ }\bibfield
  {title} {\bibinfo {title} {{Cost of counterdiabatic driving and work
  output}},\ }\href {https://doi.org/10.1103/PhysRevA.94.042132} {\bibfield
  {journal} {\bibinfo  {journal} {Phys. Rev. A}\ }\textbf {\bibinfo {volume}
  {94}},\ \bibinfo {pages} {042132} (\bibinfo {year} {2016})}\BibitemShut
  {NoStop}%
\bibitem [{\citenamefont {Torrontegui}\ \emph {et~al.}(2017)\citenamefont
  {Torrontegui}, \citenamefont {Lizuain}, \citenamefont
  {Gonz{\ifmmode\acute{a}\else\'{a}\fi}lez-Resines}, \citenamefont {Tobalina},
  \citenamefont {Ruschhaupt}, \citenamefont {Kosloff},\ and\ \citenamefont
  {Muga}}]{Torrontegui17}%
  \BibitemOpen
  \bibfield  {author} {\bibinfo {author} {\bibfnamefont {E.}~\bibnamefont
  {Torrontegui}}, \bibinfo {author} {\bibfnamefont {I.}~\bibnamefont
  {Lizuain}}, \bibinfo {author} {\bibfnamefont {S.}~\bibnamefont
  {Gonz{\ifmmode\acute{a}\else\'{a}\fi}lez-Resines}}, \bibinfo {author}
  {\bibfnamefont {A.}~\bibnamefont {Tobalina}}, \bibinfo {author}
  {\bibfnamefont {A.}~\bibnamefont {Ruschhaupt}}, \bibinfo {author}
  {\bibfnamefont {R.}~\bibnamefont {Kosloff}},\ and\ \bibinfo {author}
  {\bibfnamefont {J.~G.}\ \bibnamefont {Muga}},\ }\bibfield  {title} {\bibinfo
  {title} {{Energy consumption for shortcuts to adiabaticity}},\ }\href
  {https://doi.org/10.1103/PhysRevA.96.022133} {\bibfield  {journal} {\bibinfo
  {journal} {Phys. Rev. A}\ }\textbf {\bibinfo {volume} {96}},\ \bibinfo
  {pages} {022133} (\bibinfo {year} {2017})}\BibitemShut {NoStop}%
\bibitem [{\citenamefont {Tobalina}\ \emph {et~al.}(2019)\citenamefont
  {Tobalina}, \citenamefont {Lizuain},\ and\ \citenamefont
  {Muga}}]{Tobalina19}%
  \BibitemOpen
  \bibfield  {author} {\bibinfo {author} {\bibfnamefont {A.}~\bibnamefont
  {Tobalina}}, \bibinfo {author} {\bibfnamefont {I.}~\bibnamefont {Lizuain}},\
  and\ \bibinfo {author} {\bibfnamefont {J.~G.}\ \bibnamefont {Muga}},\
  }\bibfield  {title} {\bibinfo {title} {{Vanishing efficiency of a speeded-up
  ion-in-Paul-trap Otto engine(a)}},\ }\href
  {https://doi.org/10.1209/0295-5075/127/20005} {\bibfield  {journal} {\bibinfo
   {journal} {Europhys. Lett.}\ }\textbf {\bibinfo {volume} {127}},\ \bibinfo
  {pages} {20005} (\bibinfo {year} {2019})}\BibitemShut {NoStop}%
\bibitem [{\citenamefont {Funo}\ \emph {et~al.}(2017)\citenamefont {Funo},
  \citenamefont {Zhang}, \citenamefont {Chatou}, \citenamefont {Kim},
  \citenamefont {Ueda},\ and\ \citenamefont {del Campo}}]{Funo17}%
  \BibitemOpen
  \bibfield  {author} {\bibinfo {author} {\bibfnamefont {K.}~\bibnamefont
  {Funo}}, \bibinfo {author} {\bibfnamefont {J.-N.}\ \bibnamefont {Zhang}},
  \bibinfo {author} {\bibfnamefont {C.}~\bibnamefont {Chatou}}, \bibinfo
  {author} {\bibfnamefont {K.}~\bibnamefont {Kim}}, \bibinfo {author}
  {\bibfnamefont {M.}~\bibnamefont {Ueda}},\ and\ \bibinfo {author}
  {\bibfnamefont {A.}~\bibnamefont {del Campo}},\ }\bibfield  {title} {\bibinfo
  {title} {{Universal Work Fluctuations During Shortcuts to Adiabaticity by
  Counterdiabatic Driving}},\ }\href
  {https://doi.org/10.1103/PhysRevLett.118.100602} {\bibfield  {journal}
  {\bibinfo  {journal} {Phys. Rev. Lett.}\ }\textbf {\bibinfo {volume} {118}},\
  \bibinfo {pages} {100602} (\bibinfo {year} {2017})}\BibitemShut {NoStop}%
\bibitem [{\citenamefont {del Campo}\ \emph {et~al.}(2018)\citenamefont {del
  Campo}, \citenamefont {Chenu}, \citenamefont {Deng},\ and\ \citenamefont
  {Wu}}]{delcampo18}%
  \BibitemOpen
  \bibfield  {author} {\bibinfo {author} {\bibfnamefont {A.}~\bibnamefont {del
  Campo}}, \bibinfo {author} {\bibfnamefont {A.}~\bibnamefont {Chenu}},
  \bibinfo {author} {\bibfnamefont {S.}~\bibnamefont {Deng}},\ and\ \bibinfo
  {author} {\bibfnamefont {H.}~\bibnamefont {Wu}},\ }\bibinfo {title}
  {Friction-free quantum machines},\ in\ \href
  {https://doi.org/10.1007/978-3-319-99046-0_5} {\emph {\bibinfo {booktitle}
  {Thermodynamics in the Quantum Regime: Fundamental Aspects and New
  Directions}}},\ \bibinfo {editor} {edited by\ \bibinfo {editor}
  {\bibfnamefont {F.}~\bibnamefont {Binder}}, \bibinfo {editor} {\bibfnamefont
  {L.~A.}\ \bibnamefont {Correa}}, \bibinfo {editor} {\bibfnamefont
  {C.}~\bibnamefont {Gogolin}}, \bibinfo {editor} {\bibfnamefont
  {J.}~\bibnamefont {Anders}},\ and\ \bibinfo {editor} {\bibfnamefont
  {G.}~\bibnamefont {Adesso}}}\ (\bibinfo  {publisher} {Springer International
  Publishing},\ \bibinfo {address} {Cham},\ \bibinfo {year} {2018})\ pp.\
  \bibinfo {pages} {127--148}\BibitemShut {NoStop}%
\bibitem [{\citenamefont {Kiely}\ \emph {et~al.}(2022)\citenamefont {Kiely},
  \citenamefont {Campbell},\ and\ \citenamefont {Landi}}]{Kiely22}%
  \BibitemOpen
  \bibfield  {author} {\bibinfo {author} {\bibfnamefont {A.}~\bibnamefont
  {Kiely}}, \bibinfo {author} {\bibfnamefont {S.}~\bibnamefont {Campbell}},\
  and\ \bibinfo {author} {\bibfnamefont {G.~T.}\ \bibnamefont {Landi}},\
  }\bibfield  {title} {\bibinfo {title} {{Classical dissipative cost of quantum
  control}},\ }\href {https://doi.org/10.1103/PhysRevA.106.012202} {\bibfield
  {journal} {\bibinfo  {journal} {Phys. Rev. A}\ }\textbf {\bibinfo {volume}
  {106}},\ \bibinfo {pages} {012202} (\bibinfo {year} {2022})}\BibitemShut
  {NoStop}%
\bibitem [{\citenamefont {Carolan}\ \emph {et~al.}(2022)\citenamefont
  {Carolan}, \citenamefont {Kiely},\ and\ \citenamefont
  {Campbell}}]{Carolan22}%
  \BibitemOpen
  \bibfield  {author} {\bibinfo {author} {\bibfnamefont {E.}~\bibnamefont
  {Carolan}}, \bibinfo {author} {\bibfnamefont {A.}~\bibnamefont {Kiely}},\
  and\ \bibinfo {author} {\bibfnamefont {S.}~\bibnamefont {Campbell}},\
  }\bibfield  {title} {\bibinfo {title} {{Counterdiabatic control in the
  impulse regime}},\ }\href {https://doi.org/10.1103/PhysRevA.105.012605}
  {\bibfield  {journal} {\bibinfo  {journal} {Phys. Rev. A}\ }\textbf {\bibinfo
  {volume} {105}},\ \bibinfo {pages} {012605} (\bibinfo {year}
  {2022})}\BibitemShut {NoStop}%
\bibitem [{\citenamefont
  {Auff{\ifmmode\grave{e}\else\`{e}\fi}ves}(2022)}]{Auffeves22}%
  \BibitemOpen
  \bibfield  {author} {\bibinfo {author} {\bibfnamefont {A.}~\bibnamefont
  {Auff{\ifmmode\grave{e}\else\`{e}\fi}ves}},\ }\bibfield  {title} {\bibinfo
  {title} {{Quantum Technologies Need a Quantum Energy Initiative}},\ }\href
  {https://doi.org/10.1103/PRXQuantum.3.020101} {\bibfield  {journal} {\bibinfo
   {journal} {PRX Quantum}\ }\textbf {\bibinfo {volume} {3}},\ \bibinfo {pages}
  {020101} (\bibinfo {year} {2022})}\BibitemShut {NoStop}%
\bibitem [{\citenamefont {Fellous-Asiani}\ \emph
  {et~al.}(2023{\natexlab{a}})\citenamefont {Fellous-Asiani}, \citenamefont
  {Chai}, \citenamefont {Thonnart}, \citenamefont {Ng}, \citenamefont
  {Whitney},\ and\ \citenamefont
  {Auff{\ifmmode\grave{e}\else\`{e}\fi}ves}}]{Asiani23}%
  \BibitemOpen
  \bibfield  {author} {\bibinfo {author} {\bibfnamefont {M.}~\bibnamefont
  {Fellous-Asiani}}, \bibinfo {author} {\bibfnamefont {J.~H.}\ \bibnamefont
  {Chai}}, \bibinfo {author} {\bibfnamefont {Y.}~\bibnamefont {Thonnart}},
  \bibinfo {author} {\bibfnamefont {H.~K.}\ \bibnamefont {Ng}}, \bibinfo
  {author} {\bibfnamefont {R.~S.}\ \bibnamefont {Whitney}},\ and\ \bibinfo
  {author} {\bibfnamefont {A.}~\bibnamefont
  {Auff{\ifmmode\grave{e}\else\`{e}\fi}ves}},\ }\bibfield  {title} {\bibinfo
  {title} {{Optimizing Resource Efficiencies for Scalable Full-Stack Quantum
  Computers}},\ }\href {https://doi.org/10.1103/PRXQuantum.4.040319} {\bibfield
   {journal} {\bibinfo  {journal} {PRX Quantum}\ }\textbf {\bibinfo {volume}
  {4}},\ \bibinfo {pages} {040319} (\bibinfo {year}
  {2023}{\natexlab{a}})}\BibitemShut {NoStop}%
\bibitem [{\citenamefont {Schirmer}\ \emph {et~al.}(2004)\citenamefont
  {Schirmer}, \citenamefont {Zhang},\ and\ \citenamefont
  {Leahy}}]{schirmer:2004}%
  \BibitemOpen
  \bibfield  {author} {\bibinfo {author} {\bibfnamefont {S.~G.}\ \bibnamefont
  {Schirmer}}, \bibinfo {author} {\bibfnamefont {T.}~\bibnamefont {Zhang}},\
  and\ \bibinfo {author} {\bibfnamefont {J.~V.}\ \bibnamefont {Leahy}},\
  }\bibfield  {title} {\bibinfo {title} {Orbits of quantum states and geometry
  of bloch vectors for n-level systems},\ }\href
  {https://doi.org/10.1088/0305-4470/37/4/022} {\bibfield  {journal} {\bibinfo
  {journal} {Journal of Physics A: Mathematical and General}\ }\textbf
  {\bibinfo {volume} {37}},\ \bibinfo {pages} {1389} (\bibinfo {year}
  {2004})}\BibitemShut {NoStop}%
\bibitem [{\citenamefont {Boscain}\ \emph {et~al.}(2021)\citenamefont
  {Boscain}, \citenamefont {Sigalotti},\ and\ \citenamefont
  {Sugny}}]{Boscain21}%
  \BibitemOpen
  \bibfield  {author} {\bibinfo {author} {\bibfnamefont {U.}~\bibnamefont
  {Boscain}}, \bibinfo {author} {\bibfnamefont {M.}~\bibnamefont {Sigalotti}},\
  and\ \bibinfo {author} {\bibfnamefont {D.}~\bibnamefont {Sugny}},\ }\bibfield
   {title} {\bibinfo {title} {{Introduction to the Pontryagin Maximum Principle
  for Quantum Optimal Control}},\ }\href
  {https://doi.org/10.1103/PRXQuantum.2.030203} {\bibfield  {journal} {\bibinfo
   {journal} {PRX Quantum}\ }\textbf {\bibinfo {volume} {2}},\ \bibinfo {pages}
  {030203} (\bibinfo {year} {2021})}\BibitemShut {NoStop}%
\bibitem [{\citenamefont {Latune}(2021)}]{Latune21}%
  \BibitemOpen
  \bibfield  {author} {\bibinfo {author} {\bibfnamefont {C.~L.}\ \bibnamefont
  {Latune}},\ }\bibfield  {title} {\bibinfo {title} {{Energetic advantages of
  nonadiabatic drives combined with nonthermal quantum states}},\ }\href
  {https://doi.org/10.1103/PhysRevA.103.062221} {\bibfield  {journal} {\bibinfo
   {journal} {Phys. Rev. A}\ }\textbf {\bibinfo {volume} {103}},\ \bibinfo
  {pages} {062221} (\bibinfo {year} {2021})}\BibitemShut {NoStop}%
\bibitem [{\citenamefont {Pontryagin}\ \emph {et~al.}(1962)\citenamefont
  {Pontryagin}, \citenamefont {Boltianski}, \citenamefont {Gamkrelidze},\ and\
  \citenamefont {Mitchtchenko}}]{pontryaginbook}%
  \BibitemOpen
  \bibfield  {author} {\bibinfo {author} {\bibfnamefont {L.~S.}\ \bibnamefont
  {Pontryagin}}, \bibinfo {author} {\bibfnamefont {V.}~\bibnamefont
  {Boltianski}}, \bibinfo {author} {\bibfnamefont {R.}~\bibnamefont
  {Gamkrelidze}},\ and\ \bibinfo {author} {\bibfnamefont {E.}~\bibnamefont
  {Mitchtchenko}},\ }\href@noop {} {\emph {\bibinfo {title} {{The Mathematical
  Theory of Optimal Processes}}}}\ (\bibinfo  {publisher} {{John Wiley and
  Sons, New York}},\ \bibinfo {year} {1962})\BibitemShut {NoStop}%
\bibitem [{\citenamefont {Lee}\ and\ \citenamefont
  {Markus}(1967)}]{leemarkusbook}%
  \BibitemOpen
  \bibfield  {author} {\bibinfo {author} {\bibfnamefont {M.~M.}\ \bibnamefont
  {Lee}}\ and\ \bibinfo {author} {\bibfnamefont {L.}~\bibnamefont {Markus}},\
  }\href@noop {} {\emph {\bibinfo {title} {{Foundations of Optimal Control
  Theory}}}}\ (\bibinfo  {publisher} {{John Wiley and Sons, New York}},\
  \bibinfo {year} {1967})\BibitemShut {NoStop}%
\bibitem [{\citenamefont {Ansel}\ \emph {et~al.}(2024)\citenamefont {Ansel},
  \citenamefont {Dionis}, \citenamefont {Arrouas}, \citenamefont {Peaudecerf},
  \citenamefont {Gu\'erin}, \citenamefont {Gu\'ery-Odelin},\ and\ \citenamefont
  {Sugny}}]{Ansel24}%
  \BibitemOpen
  \bibfield  {author} {\bibinfo {author} {\bibfnamefont {Q.}~\bibnamefont
  {Ansel}}, \bibinfo {author} {\bibfnamefont {E.}~\bibnamefont {Dionis}},
  \bibinfo {author} {\bibfnamefont {F.}~\bibnamefont {Arrouas}}, \bibinfo
  {author} {\bibfnamefont {B.}~\bibnamefont {Peaudecerf}}, \bibinfo {author}
  {\bibfnamefont {S.}~\bibnamefont {Gu\'erin}}, \bibinfo {author}
  {\bibfnamefont {D.}~\bibnamefont {Gu\'ery-Odelin}},\ and\ \bibinfo {author}
  {\bibfnamefont {D.}~\bibnamefont {Sugny}},\ }\bibfield  {title} {\bibinfo
  {title} {Introduction to theoretical and experimental aspects of quantum
  optimal control},\ }\href {https://doi.org/10.1088/1361-6455/ad46a5}
  {\bibfield  {journal} {\bibinfo  {journal} {Journal of Physics B: Atomic,
  Molecular and Optical Physics}\ }\textbf {\bibinfo {volume} {57}},\ \bibinfo
  {pages} {133001} (\bibinfo {year} {2024})}\BibitemShut {NoStop}%
\bibitem [{\citenamefont {Bonnard}\ and\ \citenamefont
  {Sugny}(2012)}]{bonnard_optimal_2012}%
  \BibitemOpen
  \bibfield  {author} {\bibinfo {author} {\bibfnamefont {B.}~\bibnamefont
  {Bonnard}}\ and\ \bibinfo {author} {\bibfnamefont {D.}~\bibnamefont
  {Sugny}},\ }\href@noop {} {\emph {\bibinfo {title} {Optimal Control with
  Applications in Space and Quantum Dynamics}}},\ \bibinfo {series} {AIMS on
  applied mathematics}, Vol.~\bibinfo {volume} {5}\ (\bibinfo  {publisher}
  {American Institute of Mathematical Sciences, Springfield},\ \bibinfo {year}
  {2012})\BibitemShut {NoStop}%
\bibitem [{\citenamefont {Bason}\ \emph {et~al.}(2012)\citenamefont {Bason},
  \citenamefont {Viteau}, \citenamefont {Malossi}, \citenamefont {Huillery},
  \citenamefont {Arimondo}, \citenamefont {Ciampini}, \citenamefont {Fazio},
  \citenamefont {Giovannetti}, \citenamefont {Mannella}, ,\ and\ \citenamefont
  {Morsch}}]{bason2012}%
  \BibitemOpen
  \bibfield  {author} {\bibinfo {author} {\bibfnamefont {M.~G.}\ \bibnamefont
  {Bason}}, \bibinfo {author} {\bibfnamefont {M.}~\bibnamefont {Viteau}},
  \bibinfo {author} {\bibfnamefont {N.}~\bibnamefont {Malossi}}, \bibinfo
  {author} {\bibfnamefont {P.}~\bibnamefont {Huillery}}, \bibinfo {author}
  {\bibfnamefont {E.}~\bibnamefont {Arimondo}}, \bibinfo {author}
  {\bibfnamefont {D.}~\bibnamefont {Ciampini}}, \bibinfo {author}
  {\bibfnamefont {R.}~\bibnamefont {Fazio}}, \bibinfo {author} {\bibfnamefont
  {V.}~\bibnamefont {Giovannetti}}, \bibinfo {author} {\bibfnamefont
  {R.}~\bibnamefont {Mannella}}, ,\ and\ \bibinfo {author} {\bibfnamefont
  {O.}~\bibnamefont {Morsch}},\ }\bibfield  {title} {\bibinfo {title}
  {High-fidelity quantum driving},\ }\href@noop {} {\bibfield  {journal}
  {\bibinfo  {journal} {Nature Physics}\ }\textbf {\bibinfo {volume} {8}},\
  \bibinfo {pages} {147} (\bibinfo {year} {2012})}\BibitemShut {NoStop}%
\bibitem [{\citenamefont {Hegerfeldt}(2013)}]{hegerfeldt2013}%
  \BibitemOpen
  \bibfield  {author} {\bibinfo {author} {\bibfnamefont {G.~C.}\ \bibnamefont
  {Hegerfeldt}},\ }\bibfield  {title} {\bibinfo {title} {Driving at the quantum
  speed limit: Optimal control of a two-level system},\ }\href
  {https://doi.org/10.1103/PhysRevLett.111.260501} {\bibfield  {journal}
  {\bibinfo  {journal} {Phys. Rev. Lett.}\ }\textbf {\bibinfo {volume} {111}},\
  \bibinfo {pages} {260501} (\bibinfo {year} {2013})}\BibitemShut {NoStop}%
\bibitem [{\citenamefont {Zenesini}\ \emph {et~al.}(2009)\citenamefont
  {Zenesini}, \citenamefont {Lignier}, \citenamefont {Tayebirad}, \citenamefont
  {Radogostowicz}, \citenamefont {Ciampini}, \citenamefont {Mannella},
  \citenamefont {Wimberger}, \citenamefont {Morsch},\ and\ \citenamefont
  {Arimondo}}]{zenesini2009}%
  \BibitemOpen
  \bibfield  {author} {\bibinfo {author} {\bibfnamefont {A.}~\bibnamefont
  {Zenesini}}, \bibinfo {author} {\bibfnamefont {H.}~\bibnamefont {Lignier}},
  \bibinfo {author} {\bibfnamefont {G.}~\bibnamefont {Tayebirad}}, \bibinfo
  {author} {\bibfnamefont {J.}~\bibnamefont {Radogostowicz}}, \bibinfo {author}
  {\bibfnamefont {D.}~\bibnamefont {Ciampini}}, \bibinfo {author}
  {\bibfnamefont {R.}~\bibnamefont {Mannella}}, \bibinfo {author}
  {\bibfnamefont {S.}~\bibnamefont {Wimberger}}, \bibinfo {author}
  {\bibfnamefont {O.}~\bibnamefont {Morsch}},\ and\ \bibinfo {author}
  {\bibfnamefont {E.}~\bibnamefont {Arimondo}},\ }\bibfield  {title} {\bibinfo
  {title} {Time-resolved measurement of landau-zener tunneling in periodic
  potentials},\ }\href {https://doi.org/10.1103/PhysRevLett.103.090403}
  {\bibfield  {journal} {\bibinfo  {journal} {Phys. Rev. Lett.}\ }\textbf
  {\bibinfo {volume} {103}},\ \bibinfo {pages} {090403} (\bibinfo {year}
  {2009})}\BibitemShut {NoStop}%
\bibitem [{\citenamefont {Tayebirad}\ \emph {et~al.}(2010)\citenamefont
  {Tayebirad}, \citenamefont {Zenesini}, \citenamefont {Ciampini},
  \citenamefont {Mannella}, \citenamefont {Morsch}, \citenamefont {Arimondo},
  \citenamefont {L\"orch},\ and\ \citenamefont {Wimberger}}]{trayebirad2010}%
  \BibitemOpen
  \bibfield  {author} {\bibinfo {author} {\bibfnamefont {G.}~\bibnamefont
  {Tayebirad}}, \bibinfo {author} {\bibfnamefont {A.}~\bibnamefont {Zenesini}},
  \bibinfo {author} {\bibfnamefont {D.}~\bibnamefont {Ciampini}}, \bibinfo
  {author} {\bibfnamefont {R.}~\bibnamefont {Mannella}}, \bibinfo {author}
  {\bibfnamefont {O.}~\bibnamefont {Morsch}}, \bibinfo {author} {\bibfnamefont
  {E.}~\bibnamefont {Arimondo}}, \bibinfo {author} {\bibfnamefont
  {N.}~\bibnamefont {L\"orch}},\ and\ \bibinfo {author} {\bibfnamefont
  {S.}~\bibnamefont {Wimberger}},\ }\bibfield  {title} {\bibinfo {title}
  {Time-resolved measurement of landau-zener tunneling in different bases},\
  }\href {https://doi.org/10.1103/PhysRevA.82.013633} {\bibfield  {journal}
  {\bibinfo  {journal} {Phys. Rev. A}\ }\textbf {\bibinfo {volume} {82}},\
  \bibinfo {pages} {013633} (\bibinfo {year} {2010})}\BibitemShut {NoStop}%
\bibitem [{\citenamefont {Kobzar}\ \emph {et~al.}(2004)\citenamefont {Kobzar},
  \citenamefont {Skinner}, \citenamefont {Khaneja}, \citenamefont {Glaser},\
  and\ \citenamefont {Luy}}]{kobzar2004exploring}%
  \BibitemOpen
  \bibfield  {author} {\bibinfo {author} {\bibfnamefont {K.}~\bibnamefont
  {Kobzar}}, \bibinfo {author} {\bibfnamefont {T.~E.}\ \bibnamefont {Skinner}},
  \bibinfo {author} {\bibfnamefont {N.}~\bibnamefont {Khaneja}}, \bibinfo
  {author} {\bibfnamefont {S.~J.}\ \bibnamefont {Glaser}},\ and\ \bibinfo
  {author} {\bibfnamefont {B.}~\bibnamefont {Luy}},\ }\bibfield  {title}
  {\bibinfo {title} {Exploring the limits of broadband excitation and inversion
  pulses},\ }\href@noop {} {\bibfield  {journal} {\bibinfo  {journal} {Journal
  of Magnetic Resonance}\ }\textbf {\bibinfo {volume} {170}},\ \bibinfo {pages}
  {236} (\bibinfo {year} {2004})}\BibitemShut {NoStop}%
\bibitem [{\citenamefont {Kobzar}\ \emph {et~al.}(2012)\citenamefont {Kobzar},
  \citenamefont {Ehni}, \citenamefont {Skinner}, \citenamefont {Glaser},\ and\
  \citenamefont {Luy}}]{kobzar2012exploring}%
  \BibitemOpen
  \bibfield  {author} {\bibinfo {author} {\bibfnamefont {K.}~\bibnamefont
  {Kobzar}}, \bibinfo {author} {\bibfnamefont {S.}~\bibnamefont {Ehni}},
  \bibinfo {author} {\bibfnamefont {T.~E.}\ \bibnamefont {Skinner}}, \bibinfo
  {author} {\bibfnamefont {S.~J.}\ \bibnamefont {Glaser}},\ and\ \bibinfo
  {author} {\bibfnamefont {B.}~\bibnamefont {Luy}},\ }\bibfield  {title}
  {\bibinfo {title} {Exploring the limits of broadband 90 and 180 universal
  rotation pulses},\ }\href@noop {} {\bibfield  {journal} {\bibinfo  {journal}
  {Journal of Magnetic Resonance}\ }\textbf {\bibinfo {volume} {225}},\
  \bibinfo {pages} {142} (\bibinfo {year} {2012})}\BibitemShut {NoStop}%
\bibitem [{\citenamefont {Khaneja}\ \emph {et~al.}(2005)\citenamefont
  {Khaneja}, \citenamefont {Reiss}, \citenamefont {Kehlet}, \citenamefont
  {Schulte-Herbrüggen},\ and\ \citenamefont {Glaser}}]{khaneja_optimal_2005}%
  \BibitemOpen
  \bibfield  {author} {\bibinfo {author} {\bibfnamefont {N.}~\bibnamefont
  {Khaneja}}, \bibinfo {author} {\bibfnamefont {T.}~\bibnamefont {Reiss}},
  \bibinfo {author} {\bibfnamefont {C.}~\bibnamefont {Kehlet}}, \bibinfo
  {author} {\bibfnamefont {T.}~\bibnamefont {Schulte-Herbrüggen}},\ and\
  \bibinfo {author} {\bibfnamefont {S.~J.}\ \bibnamefont {Glaser}},\ }\bibfield
   {title} {\bibinfo {title} {Optimal control of coupled spin dynamics: design
  of {NMR} pulse sequences by gradient ascent algorithms},\ }\href
  {https://doi.org/10.1016/j.jmr.2004.11.004} {\bibfield  {journal} {\bibinfo
  {journal} {J. Magn. Res.}\ }\textbf {\bibinfo {volume} {172}},\ \bibinfo
  {pages} {296} (\bibinfo {year} {2005})}\BibitemShut {NoStop}%
\bibitem [{\citenamefont {Chen}\ \emph {et~al.}(2010)\citenamefont {Chen},
  \citenamefont {Lizuain}, \citenamefont {Ruschhaupt}, \citenamefont
  {Gu\'ery-Odelin},\ and\ \citenamefont {Muga}}]{Chen:2010}%
  \BibitemOpen
  \bibfield  {author} {\bibinfo {author} {\bibfnamefont {X.}~\bibnamefont
  {Chen}}, \bibinfo {author} {\bibfnamefont {I.}~\bibnamefont {Lizuain}},
  \bibinfo {author} {\bibfnamefont {A.}~\bibnamefont {Ruschhaupt}}, \bibinfo
  {author} {\bibfnamefont {D.}~\bibnamefont {Gu\'ery-Odelin}},\ and\ \bibinfo
  {author} {\bibfnamefont {J.~G.}\ \bibnamefont {Muga}},\ }\bibfield  {title}
  {\bibinfo {title} {Shortcut to adiabatic passage in two- and three-level
  atoms},\ }\href {https://doi.org/10.1103/PhysRevLett.105.123003} {\bibfield
  {journal} {\bibinfo  {journal} {Phys. Rev. Lett.}\ }\textbf {\bibinfo
  {volume} {105}},\ \bibinfo {pages} {123003} (\bibinfo {year}
  {2010})}\BibitemShut {NoStop}%
\bibitem [{\citenamefont {del Campo}(2021)}]{delcampo21}%
  \BibitemOpen
  \bibfield  {author} {\bibinfo {author} {\bibfnamefont {A.}~\bibnamefont {del
  Campo}},\ }\bibfield  {title} {\bibinfo {title} {{Probing Quantum Speed
  Limits with Ultracold Gases}},\ }\href
  {https://doi.org/10.1103/PhysRevLett.126.180603} {\bibfield  {journal}
  {\bibinfo  {journal} {Phys. Rev. Lett.}\ }\textbf {\bibinfo {volume} {126}},\
  \bibinfo {pages} {180603} (\bibinfo {year} {2021})}\BibitemShut {NoStop}%
\bibitem [{\citenamefont {del Campo}\ \emph {et~al.}(2013)\citenamefont {del
  Campo}, \citenamefont {Egusquiza}, \citenamefont {Plenio},\ and\
  \citenamefont {Huelga}}]{delCampo2013Jan}%
  \BibitemOpen
  \bibfield  {author} {\bibinfo {author} {\bibfnamefont {A.}~\bibnamefont {del
  Campo}}, \bibinfo {author} {\bibfnamefont {I.~L.}\ \bibnamefont {Egusquiza}},
  \bibinfo {author} {\bibfnamefont {M.~B.}\ \bibnamefont {Plenio}},\ and\
  \bibinfo {author} {\bibfnamefont {S.~F.}\ \bibnamefont {Huelga}},\ }\bibfield
   {title} {\bibinfo {title} {{Quantum Speed Limits in Open System Dynamics}},\
  }\href {https://doi.org/10.1103/PhysRevLett.110.050403} {\bibfield  {journal}
  {\bibinfo  {journal} {Phys. Rev. Lett.}\ }\textbf {\bibinfo {volume} {110}},\
  \bibinfo {pages} {050403} (\bibinfo {year} {2013})}\BibitemShut {NoStop}%
\bibitem [{\citenamefont {Koch}(2016)}]{Koch2016}%
  \BibitemOpen
  \bibfield  {author} {\bibinfo {author} {\bibfnamefont {C.~P.}\ \bibnamefont
  {Koch}},\ }\bibfield  {title} {\bibinfo {title} {{Controlling open quantum
  systems: tools, achievements, and limitations}},\ }\href
  {https://doi.org/10.1088/0953-8984/28/21/213001} {\bibfield  {journal}
  {\bibinfo  {journal} {J. Phys.: Condens. Matter}\ }\textbf {\bibinfo {volume}
  {28}},\ \bibinfo {pages} {213001} (\bibinfo {year} {2016})}\BibitemShut
  {NoStop}%
\bibitem [{\citenamefont {Alipour}\ \emph {et~al.}(2020)\citenamefont
  {Alipour}, \citenamefont {Chenu}, \citenamefont {Rezakhani},\ and\
  \citenamefont {del Campo}}]{Alipour2020}%
  \BibitemOpen
  \bibfield  {author} {\bibinfo {author} {\bibfnamefont {S.}~\bibnamefont
  {Alipour}}, \bibinfo {author} {\bibfnamefont {A.}~\bibnamefont {Chenu}},
  \bibinfo {author} {\bibfnamefont {A.~T.}\ \bibnamefont {Rezakhani}},\ and\
  \bibinfo {author} {\bibfnamefont {A.}~\bibnamefont {del Campo}},\ }\bibfield
  {title} {\bibinfo {title} {{Shortcuts to Adiabaticity in Driven Open Quantum
  Systems: Balanced Gain and Loss and Non-Markovian Evolution}},\ }\href
  {https://doi.org/10.22331/q-2020-09-28-336} {\bibfield  {journal} {\bibinfo
  {journal} {Quantum}\ }\textbf {\bibinfo {volume} {4}},\ \bibinfo {pages}
  {336} (\bibinfo {year} {2020})},\ \Eprint
  {https://arxiv.org/abs/1907.07460v2} {1907.07460v2} \BibitemShut {NoStop}%
\bibitem [{\citenamefont {Fellous-Asiani}\ \emph
  {et~al.}(2023{\natexlab{b}})\citenamefont {Fellous-Asiani}, \citenamefont
  {Chai}, \citenamefont {Thonnart}, \citenamefont {Ng}, \citenamefont
  {Whitney},\ and\ \citenamefont
  {Auff{\ifmmode\grave{e}\else\`{e}\fi}ves}}]{Fellous-Asiani23}%
  \BibitemOpen
  \bibfield  {author} {\bibinfo {author} {\bibfnamefont {M.}~\bibnamefont
  {Fellous-Asiani}}, \bibinfo {author} {\bibfnamefont {J.~H.}\ \bibnamefont
  {Chai}}, \bibinfo {author} {\bibfnamefont {Y.}~\bibnamefont {Thonnart}},
  \bibinfo {author} {\bibfnamefont {H.~K.}\ \bibnamefont {Ng}}, \bibinfo
  {author} {\bibfnamefont {R.~S.}\ \bibnamefont {Whitney}},\ and\ \bibinfo
  {author} {\bibfnamefont {A.}~\bibnamefont
  {Auff{\ifmmode\grave{e}\else\`{e}\fi}ves}},\ }\bibfield  {title} {\bibinfo
  {title} {{Optimizing Resource Efficiencies for Scalable Full-Stack Quantum
  Computers}},\ }\href {https://doi.org/10.1103/PRXQuantum.4.040319} {\bibfield
   {journal} {\bibinfo  {journal} {PRX Quantum}\ }\textbf {\bibinfo {volume}
  {4}},\ \bibinfo {pages} {040319} (\bibinfo {year}
  {2023}{\natexlab{b}})}\BibitemShut {NoStop}%
\end{thebibliography}%

\end{document}